\documentclass[journal]{IEEEtran}

\usepackage{url}
\makeatletter
\g@addto@macro{\UrlBreaks}{\UrlOrds}
\makeatother
\usepackage{blindtext, graphicx}
\usepackage{qtree}
\usepackage[table]{xcolor}
\usepackage{url}
\usepackage{tabularx}
\usepackage{cite}
\definecolor{green}{HTML}{66FF66}
\definecolor{myGreen}{HTML}{009900}
\usepackage{fixltx2e}
\usepackage{epstopdf}
\usepackage{amsmath}
\usepackage{algorithmic}
\usepackage{array}
\usepackage{mdwmath}
\usepackage{mdwtab}
\usepackage[nolist]{acronym}
\usepackage{multirow}
\usepackage{graphicx}
\usepackage{float}
\usepackage{epsfig,amssymb}
\usepackage{epstopdf}
\usepackage{flushend}
\usepackage{adjustbox}
\usepackage{epsfig}
\usepackage{eurosym}
\usepackage{epsfig,amssymb}
\usepackage{epstopdf}
\usepackage{caption}
\usepackage{subcaption}
\usepackage{lipsum}
\usepackage{authblk}
\usepackage{amsfonts} 

\usepackage{hyperref}
\usepackage{graphicx}
\usepackage{xcolor}
\definecolor{darkblue}{rgb}{0,0,0.5}
\usepackage{transparent}
\usepackage{filecontents}

\IEEEoverridecommandlockouts

\begin{document}
\title{From 4G to 5G: Self-organized Network Management meets Machine Learning}
\author[1]{Jessica Moysen}
\author[2]{Lorenza Giupponi}
\affil[1]{Department of Signal and Theory Communications\\
Universitat Polit\`{e}cnica de Catalunya-UPC\authorcr Email: {\tt jessica.moysen@tsc.upc.edu}\vspace{1.5ex}}
\affil[2]{Communications Network Division\\
Centre Tecnol\`{o}gic de Telecomunicacions de Catalunya-CTTC\authorcr Email:
  {\tt lorenza.giupponi@cttc.es} \vspace{-2ex}

\thanks {The research leading to these results has received funding from the Spanish Ministry of Economy and Competitiveness under grant TEC2014-60491-R (Project 5GNORM). This work also was supported by the Spanish National Science Council and ERFD funds under Project TEC2014-60258-C2-2-R.
}}

\renewcommand\Authands{ and }

 \begin{acronym}[AAAAAAA]
\acro{AAS}[AAS]{Active Antenna Systems}
\acro{AC}[AC]{Actor Critic}
\acro{AI}[AI]{ Artificial Intelligence}
\acro{ABS}[ABS]{Almost Blank Subframes}
\acro{AIP}[AIP]{Administrative Incentive Pricing}
\acro{ANA}[ANA]{Autonomic Network Architecture}
\acro{ANR}[ANR]{Automatic Neighbour Relation}
\acro{ANN}[ANN]{Artificial Neural Network}
\acro{AP}[AP]{Access Point}
\acro{API}[API]{Application Programming Interface}
\acro{BeFemto}[BeFemto]{Broadband Evolved FEMTO Networks}
\acro{BER}[BER]{Bit Error Rate}
\acro{BLER}[BLER]{Block Error Rate}
\acro{BIONETS}[BIONETS]{Genetically inspired networks}
\acro{2TBN}[2TBN]{two-stage temporal Bayesian network}
\acro{BS}[BS]{Base Station}
\acro{CESC}[CESC]{Cloud-Enabled Small Cell}
\acro{COGEU}[COGEU]{Cognitive radio systems for efficient sharing of TV white spaces in EUropean context}
\acro{CAPEX}[CAPEX]{Capital Expenditure}
\acro{CASCADAS}[CASCADAS]{Component-ware for Autonomic, Situation-aware Communications, and Dynamically Adaptable Services}
\acro{CATNETS}[CATNETS]{Evaluation of the Catallaxy Paradigm for Decentralized Operation of Dynamic Application Networks}
\acro{CET}[CET]{Changes Electrical Tilt}
\acro{CG}[CG]{Coordination Game}
\acro{C-SON}[C-SON]{Centralized SON}
\acro{CIO}[CIO]{Cell Individual Offset}
\acro{CCO}[CCO]{Coverage and Capacity Optimization}
\acro{COR}[COR]{Cell Outage Recovery}
\acro{CDF}[CDF]{Cumulative Distribution Function}
\acro{CDR}[CDR]{Charging Data Records}
\acro{COC}[COC]{Cell Outage Compensation}
\acro{COD}[COD]{Cell Outage Detection}
\acro{COM}[COM]{Cell Outage Management}
\acro{CoMP}[CoMP]{Coordinated Multi Points}
\acro{COOPCOM}[COOPCOM]{Comunicaciones Cooperativas y Oportunistas en Redes Inal\'{a}mbricas}
\acro{COGNET}[COGNET]{Cognitive networks}
\acro{CRF}[CRF]{Conditional Random Field}
\acro{CRS}[CRS]{Cognitive Radio System}
\acro{CQI}[CQI]{Channel Quality Indicator}
\acro{CTTC}[CTTC]{Centre Tecnològic de Telecomunicacions de Catalunya}
\acro{DA}[DA]{Discriminant Analysis}
\acro{D-SON}[D-SON]{Distributed SON}
\acro{DCI}[DCI]{Data Control Indication}
\acro{DBM}[DBM]{Deep Boltzmann Machine}
\acro{DBN}[DBN]{Deep Belief Network}
\acro{DNN}[DNN]{Deep Neural Network}
\acro{DL}[DL]{Downlink}
\acro{DP}[DP]{Dynamic Programming}
\acro{DSA}[DSA]{Dynamic Spectrum Access}
\acro{DT}[DT]{Decision Trees}
\acro{E3}[E3]{End-to-End Efficiency}
\acro{eNB}[eNB]{Enhanced Node Base station}
\acro{eNBs}[eNBs]{Enhanced Node Base stations}
\acro{EIRP}[EIRP]{Equivalent Isotropically Radiated Power}
\acro{EPC}[EPC]{Evolved Packet Core}
\acro{ETRI}[ETRI]{Electronics and Telecomunications Research Institute}
\acro{ES}[ES]{Energy Saving}
\acro{E-UTRAN}[E-UTRAN]{Evolved Universal Terrestrial Radio Access Network}
\acro{ETSI}[ETSI]{European Telecommunications Standards Institute}
\acro{FFR}[FFR]{Fractional Frequency Reuse}
\acro{FL}[FL]{Fuzzy Logic}
\acro{FE}[FE]{Feature Extraction}
\acro{FS}[FS]{Feature Selection}
\acro{GT}[GT]{Game Theory}
\acro{Gandalf}[Gandalf]{Monitoring and Self-tuning of RRM parameters in a Multi-System Network}
\acro{GLM}[GLM]{Generalized Linear Models}
\acro{GPI}[GPI]{Generalized Policy Iteration}
\acro{GPRS}[GPRS]{General Packet Radio Service}
\acro{GSM}[GSM]{Global System for Mobile Communications}
\acro{3GPP}[3GPP]{3rd Generation Partnership Project}
\acro{5GPPP}[5GPPP]{5G Infrastructure Public Private Partnership}
\acro{GPRS}[GPRS]{General Packet Radio Service}
\acro{GSM}[GSM]{General System for Mobile Communications}
\acro{HAGGEL}[HAGGEL]{An Innovative Paradigm for Autonomic Opportunistic Communication}
\acro{HeNB}[HeNB]{Home eNodeB}
\acro{Het-Net}[Het-Net]{Heterogenous Network}
\acro{HMM}[HMM]{Hidden Markov Model}
\acro{HO}[HO]{Handover}
\acro{HII}[HII]{High Interference Indicator}
\acro{IRP}[IRP]{Integration Reference Point}
\acro{IS}[IS]{Information Service}
\acro{IEEE}[IEEE]{Institute of Electrical and Electronics Engineers}
\acro{ICIC}[ICIC]{Inter-Cell Interference Coordination}
\acro{IMS}[IMS]{IP Multimedia Subsystem}
\acro{IoT}[IoT]{Internet of Things}
\acro{IRAT}[IRAT]{Inter-Radio Access Technology}
\acro{$k$-NN}[$k$-NN]{$k$-Nearest Neighbors}
\acro{KPI}[KPI]{Key Performance Indicator}
\acro{LENA}[LENA]{LTE-EPC Network Simulator}
\acro{LTE}[LTE]{Long Term Evolution}
\acro{LTE-Advanced}[LTE-Advanced]{Long Term Evolution Advanced}
\acro{LB}[LB]{Load Balancing}
\acro{LTE-U}[LTE-U]{LTE-Unlicensed}
\acro{LAA}[LAA]{Licensed Assisted Access}
\acro{MAC}[MAC]{Media Access Control}
\acro{MDT}[MDT]{Minimization of Drive Tests}
\acro{M2M}[M2M]{Machine to Machine}
\acro{MIMO}[MIMO]{Multiple-input Multiple-output}
\acro{MC}[MC]{Monte Carlo}
\acro{MCS}[MCS]{Modulation and Coding Scheme}
\acro{MDP}[MDP]{Markov Decision Process}
\acro{MONOTAS}[MONOTAS]{Mobile Network Optimisation Through Advanced Simulation}
\acro{MCC}[MCC]{Mobile Cloud Computing}
\acro{MLB}[MLB]{Mobility Load Balancing}
\acro{ML}[ML]{Machine Learning}
\acro{MLL}[ML]{Maximum Likelihood}
\acro{MRO}[MRO]{Mobility Robustness$/$Handover Optimisation}
\acro{MWC}[MWC]{Mobile World Congress}
\acro{NB}[NB]{Naive Bayes}
\acro{NBs}[NBs]{Node Base station}
\acro{NE}[NE]{Network Element}
\acro{NET-REFOUND}[NET-REFOUND]{Network research foundations}
\acro{NGMN}[NGMN]{Next Generation Mobile Networks}
\acro{NMS}[NMS]{Network Management Systems}
\acro{NM}[NM]{Network Management}
\acro{NRM}[NRM]{Network Resource Model}
\acro{NFV}[NFV]{Network Functions Virtualisation}
\acro{OAM}[OAM]{Operations, Administration, and Maintenance}
\acro{OFDMA}[OFDMA]{Orthogonal Frequency Division Multiple Access}
\acro{OI}[OI]{Overload Indicator}
\acro{OMC}[OMC]{Operation and Maintenance Center}
\acro{OSS}[OSS]{Operation and Support System}
\acro{OPEX}[OPEX]{Operational Expenditure}
\acro{OMC}[OMC]{Operation and Maintenance Center}
\acro{PCI}[PCI]{Physical Cell ID}
\acro{PC}[PC]{Principal Component}
\acro{Pc}[Pc]{Power control}
\acro{PCA}[PCA]{Principal component analysis}
\acro{PCI}[PCI]{Automated Configuration of Physical Cell Identity}
\acro{PItoRC}[PItoRC]{Policy Iteration to Resource Conflicts}
\acro{PDF}[PDF]{Probability Density Function}
\acro{PDSCH}[PDSCH]{Physical Downlink Shared Channel}
\acro{PDU}[PDU]{Protocol Data Unit}
\acro{PDSCH}[PDSCH]{Physical Downlink Shared Channel}
\acro{PUSCH}[PUSCH]{Physical Uplink Shared Channel}
\acro{PDCCH}[PDCCH]{Physical Downlink Control Channel}
\acro{PGW}[PGW]{PDN Gateway}
\acro{SGW}[SGW]{Serving Gateway}
\acro{PF}[PF]{Proportional Fair Scheduler}
\acro{PHR}[PHR]{Power Headroom Report}
\acro{PM}[PM]{Performance Management}
\acro{PRB}[PRB]{Physical Resource Block}
\acro{PS}[PS]{Packet Switched}
\acro{PSD}[PSD]{Power Spectral Density}
\acro{PU}[PU]{Primary User}
\acro{QoS}[QoS]{Quality of Service}
\acro{QoE}[QoE]{Quality of Experience}
\acro{QAM}[QAM]{Quadrature Amplitude Modulation}
\acro{RBF}[RBF]{Radial basis function}
\acro{RBM}[RBM]{Restricted Botlzmann Machine}
\acro{RACH}[RACH]{Random Access Channel}
\acro{RAN}[RAN]{Radio Access Network}
\acro{RAT}[RAT]{Radio Access Technologies}
\acro{RET}[RET]{Remote Electrical Tilt}
\acro{RF}[RF]{Random Forest}
\acro{RB}[RB]{Resource Block}
\acro{RBG}[RBG]{Resource Block Group}
\acro{RBG}[RBG]{Restricted Boltzmann Machine}
\acro{REM}[REM]{Radio Environment Map}
\acro{RL}[RL]{Reinforcement Learning}
\acro{RLF}[RLF]{Radio Link Failure}
\acro{RNTP}[RNTP]{Relative Narrowband Transmit Power}
\acro{RLC}[RLC]{Radio Link Control}
\acro{RRC}[RRC]{Radio Resource Control}
\acro{RRM}[RRM]{Radio Resource Management}
\acro{RS}[RS]{Reference Signal}
\acro{RSRP}[RSRP]{Reference Symbol Received Power}
\acro{RSRQ}[RSRQ]{Reference Symbol Received Quality}
\acro{SAC}[SAC]{Situated Autonomic Communications}
\acro{SDSE}[SDN]{Self-Organized Network}
\acro{SDSE}[SDSE]{Strongly Dominant Strategy Equilibrium}
\acro{SELFNET}[SELFNET]{Framework for Self-Organized Network Management in virtualized
and Software Defined Networks}
\acro{SEMAFOUR}[SEMAFOUR]{Self-Management for Unified Heterogeneous Radio Access Networks}
\acro{SESAME}[SESAME]{Small cell coordination for multi-tenancy and edge services}
\acro{SG}[SG]{Stochastic Games}
\acro{SGSN}[SGSN]{Serving GPRS Support Node}
\acro{SH}[SH]{Self Healing}
\acro{SINR}[SINR]{Signal to Interference and Noise Ratio}
\acro{SLA}[SLA]{Service Level Agreement}
\acro{SM}[SM]{Saturation Mode}
\acro{SML}[SML]{Stochastic Maximum Likelihood}
\acro{SPCA}[SPCA]{Sparse Principal Component Analysis}
\acro{SVM}[SVMs]{Support Vector Machines}
\acro{SVR}[SVR]{Support Vector Regression}
\acro{SDN}[SDN]{Software Defined Network}
\acro{TCE}[TCE]{Trace Collection Entity}
\acro{SL}[SL]{Supervised Learning}
\acro{SO}[SO]{Self organization}
\acro{SOCRATES}[SOCRATES]{Self-Optimisation and self-ConfiguRATion in wirelEss networkS}
\acro{SOFOCLES}[SOFOCLES]{Self-organized FemtOCeLls for broadband sErviceS}
\acro{SOM}[SOM]{Self-organizing Map}
\acro{SON}[SON]{Self-organizing Network}
\acro{SOS}[SOS]{Self organized System}
\acro{SS}[SS]{Solution Sets}
\acro{SL}[SL]{Supervised Learning}
\acro{SIMO}[SIMO]{Single-input Multiple-output}
\acro{subMDP}[subMDP]{Markov decision sub-process}
\acro{TA}[TA]{Timing Advance}
\acro{TCE}[TCE]{Trace Collection Entity}
\acro{TCP}[TCP]{Transmission Control Protocol}
\acro{TD}[TD]{Temporal Difference}
\acro{TTT}[TTT]{Time to Trigger}
\acro{TTI}[TTI]{Transmission Time Interval}
\acro{TBS}[TBS]{Transport Block Size}
\acro{TXP}[TXP]{Transmission Power}
\acro{UDN}[UDN]{Ultra ­Dense Network}
\acro{UE}[UE]{User Equipment}
\acro{UEs}[UEs]{User Equipments}
\acro{UMTS}[UMTS]{Universal Mobile Telecommunications System}
\acro{UL}[UL]{Uplink}
\acro{UL}[UL]{Unsupervised Learning}
\acro{UTRAN}[UTRAN]{Universal Terrestrial Radio Access Network}
\acro{WLAN}[WLAN]{Wireless Local Area Network}
\acro{WCQI}[WCQI]{wideband CQI}

\end{acronym} 
\maketitle
\begin{abstract}
In this paper, we provide an analysis of self-organized network management, with an end-to-end perspective of the network. Self-organization as applied to cellular networks is usually referred to \acp{SON}, and it is a key driver for improving \ac{OAM} activities. SON aims at reducing the cost of installation and management of 4G and future 5G networks, by simplifying operational tasks through the capability to configure, optimize and heal itself. To satisfy 5G network management requirements, this autonomous management vision has to be extended to the end to end network.
In literature and also in some instances of products available in the market, \ac{ML} has been identified as the key tool to implement autonomous adaptability and take advantage of experience when making decisions. In this paper, we survey how network management can significantly benefit from ML solutions. We review and provide the basic concepts and taxonomy for SON, network management and ML. We analyse the available state of the art in the literature, standardization, and in the market. We pay special attention to \ac{3GPP} evolution in the area of network management and to the data that can be extracted from \ac{3GPP} networks, in order to gain knowledge and experience in how the network is working, and improve network performance in a proactive way. Finally, we go through the main challenges associated with this line of research, in both 4G and in what 5G is getting designed, while identifying new directions for research.

\end{abstract}
\IEEEpeerreviewmaketitle

\section{Introduction}
\label{sec:intro}
Traditionally, and up to 4G, the evolution from one generation of mobile networks to another, has been driven by hardware technology advancements. The revolution of 5G is different, and novel advancements of software technology will be critical, especially in the way the network will be managed. With the advent of these software advancements, and unprecedented levels of computational capacity, the vision of autonomous network management can be put into practice taking advantage of also other cross-disciplinary knowledge advancements in the area of Machine Learning. This vision aligns with the concepts of self-awareness, self-configuration, self-optimization, and self-healing, which have already been defined in the area of network management. We give an special emphasis to the access segment of 4G cellular \ac{LTE} network through the concept of \acf{SON}. \ac{SON} is a common term used to refer to mobile network automation and minimization of human intervention in the cellular/wireless network management. This concept has been introduced by \ac{3GPP} in Release 8 and it has been expanding across subsequent releases. \ac{3GPP} work is inspired by a set of requirements defined by the operators' Alliance \ac{NGMN}. The main objective of SON can be roughly classified into three main points: 1) to bring intelligence and autonomous adaptability into cellular networks, 2) to reduce capital and operation expenditures (CAPEX/OPEX), 3) to enhance network performances in terms of network capacity, coverage, offered service/experience, etc. SON is considered today as a driving technology that aims at improving spectral efficiency, simplifying management, and reducing the operation costs of the next generation \acp{RAN}. The overall complex SON problem has been decomposed in a set of useful use cases, which have been identified by \ac{3GPP}, \ac{NGMN}, \ac{5GPPP} and different EU projects \cite{Gandalf,SOCRATES,SEMAFOUR,SESAME,SELFNET,COGNET}.
The academic literature has dedicated significant effort to SON algorithms in the context of the above mentioned use cases, providing smart solutions to optimize network operator performance, expenses and users' experience. Many of these works are already reviewed here \cite{survey}. The market also offers already complete sets of SON solutions, (e.g. \cite{sistelbanda,qualcom,huawei,airhop}, among others) many products have been advertised and presented in \ac{MWC} 2016 and 2017 \cite{cellwize, aviat}. For example, AirHop's eSON from Jio $\&$ AirHop communications \cite{airhop}, which employs a multi-vendor, multi-technology, real-time SON solution based on scalable and virtualized software platform has recently been awarded for the 2016 Small Cell Forum \ac{Het-Net} management software and service award \cite{smallcellForum}.

However, to the best of our knowledge, the SON solutions available in the market are 1) mainly based on heuristics, 2) the automated information processing is usually limited to low complexity solutions like triggering, 3) many operations are still done manually (e.g. network faults are usually fixed directly by engineers), 4) SON solutions do not really capitalize on the huge amount of information that is available in mobile networks to build next generation network management solutions, and 5) several open challenges are still unsolved, like the problem of coordination of SON functions \cite{3GPPwork, challengesSON}, or the proper solution of the trade-off between centralized and distributed SON implementations \cite{nsn}. In addition, this self-organized network management vision should be extended also beyond the \ac{RAN} segment and should include all the segments of the network, from the access to the core, while fulfilling the requirements of different kind of vertical service instances. To achieve this vision, the networking world is exploring new directions. \ac{NFV} is expected to bring the economy of scale of the Information Technology industry to the Telecom industry. When combining \ac{NFV} with \ac{SDN} principles, the benefits of programmability and flexibility are brought to the fore.

Another aspect that should be considered is that, as we observed in  \cite{bDemp} a huge amount of data is currently already generated in 4G networks  during normal operations by control and management functions, and more is expected to come in 5G networks due to the densification process \cite{densification}, heterogeneity in layers and technologies, the additional control and management complexity in \ac{NFV} and \ac{SDN} architectures, the advent of \ac{M2M} and \ac{IoT} paradigms, the increasing variety of application and services, each with distinct traffic patterns and \ac{QoS}/\ac{QoE} requirements, etc.
5G network management is expected to provide a whole new set of challenges due to 1) the need to manage future network complexity, due to ultra dense deployments, heterogeneous nodes, networks, applications, \acp{RAN} coexisting in the same setting, 2) the need to manage very dynamic networks, where operators may do not have any control in the deployment of some nodes (e.g. femto-cells), energy saving policies are in place generating a fluctuating number of nodes, active antennas are a reality, etc. 3) the need to support $1000$x traffic, and $10$x users, and improve energy efficiency, 4) the need to improve the experience of the users by enabling Gbps speeds, and highly reduced latency, 5) the need to manage new virtualized architectures, 6) the need to handle heterogeneous spectrum access privileges through novel \ac{LTE-U}, \ac{LAA}, MuLTEFire paradigms and the availability of both traditional sub - 6 GHz bands, and above 6 GHz mmWave bands.
In this challenging context, we believe that the use of SON and of smart network management policies is crucial and inevitable for operators running multi-RAT, multi-vendor, multi-layer networks, where an overwhelming number of parameters need to be configured and optimized. The high level objective is to make the networks 1) more self-aware, by exploiting the information already available in the network to gain experience in the network management, 2) more self-adaptive, by exploiting intelligent control decisions procedures which allow to automatize the decision processes based on the experience.

We believe that \acf{ML} can be effectively used to allow the network to learn from experience, while improving performance. In particular, big data analytics, through the analysis of data already generated by the network, can pursue the self-awareness by driving the network management from reactive to predictive. Big data analytics are currently receiving big attention in research and in the market, due to their capability of providing insightful information from the analysis of data already available to operators.

In this paper, we will not focus on these uses of data analytics and ML, but we will only focus on the applications to 4/5G network management. Differently from other surveys on SON proposed before \cite{survey, dressler} or from other surveys related with 5G network management \cite{SDN-survey}, we focus here on the study and analysis of the available literature on SON and network management considering \ac{ML} as the tool to implement automation and self-organization, from a 5G perspective. We review and provide the basic concepts and taxonomy of traditional SON and 5G network management in Section \ref{sec:son}. We pay special attention to the evolution of \ac{3GPP} in the area, following its nomenclature, and referring to the specific use cases defined by the standard in this matters. We then provide, in Section \ref{sec:ml}, guidelines to select the most appropriate ML algorithm and approach, based on the network management issue to address. In Section \ref{sec:MLinSON}, we review the main sources of information
relevant for a knowledge based network management, data that is actually already generated by the network, and we survey the literature on \ac{ML}-based network management. We then highlight challenges for future works in Section \ref{sec:challenges}. Finally, Section \ref{sec:conclusions} concludes the survey.

\section{Self Organizing Network (SON) and Network Management}
\label{sec:son}

SON is a key driver to maximize total performance in cellular networks. The main idea is to bring into them intelligence and autonomous adaptability by diminishing human involvement, while enhancing network performance, in terms of network capacity, coverage and service quality. It aims at reducing the cost of installation and management by simplifying operational tasks through the capability to configure, optimize and heal itself.

The main motivation behind the increasing interest in the introduction of SON from operators, standardization bodies and projects is twofold. On the one hand, from the market perspective, the ever increasing demand for a diversity of offered services, and the need to reduce the time to market of innovative services, further add to the pressure to remain competitive by effectuating cost reductions. On the other hand, from the technical perspective, the complexity and large scale of future radio access technologies imposes significant operational challenges due to the multitude of tunable parameters and the intricate dependencies among them. In addition, the advent of heterogeneous networks is expected to tremendously increase the number of nodes in this new ecosystem, so classic manual and field trial design approaches are just impractical.

Similarly, manual optimization processes or fault diagnosis and cure, performed by experts are no longer efficient and need to be automatized, as this causes time intensive experiments with limited operational scope, or delayed, manual and poor handling of cell$/$sites failures. Key operational tasks, such as radio network planning and optimization are largely separated nowadays and this causes intrinsic shortcomings like the abstraction of access technologies for network planning purposes, or the consideration of performance indicators that are of limited relevance to the end user's service perception. These problems have been approached through SON by European projects such as SOCRATES \cite{SOCRATES}, and Gandalf \cite{Gandalf}. Also FP7 and \ac{5GPPP} EU projects have been dealing with SON. In particular FP7 SEMAFOUR \cite{SEMAFOUR}, which develops a unified self-management to operate complex HetNets. Among 5GPPP projects, we highlight SESAME \cite{SESAME}, which proposes the \ac{CESC} concept, i.e., a new multi-operator enabled Small Cell by deploying \acf{NFV}, supporting powerful self-management inside the access network infrastructure. In terms of self-organized network management in \ac{SDN} and \ac{NFV}, the article project aims at enabling the use of ML to achieve real time autonomous 5G network management \cite{SELFNET}. In particular, the project explores a smart integration of state-of-the-art technologies in \ac{SDN}, \ac{NFV}), \ac{SON}, Cloud computing, Artificial intelligence. The COGNET project \cite{COGNET} has similar objectives and aims at developing several operators' use cases by applying ML algorithms.
\begin{figure}[t!]
\centering

\includegraphics[width=\linewidth]{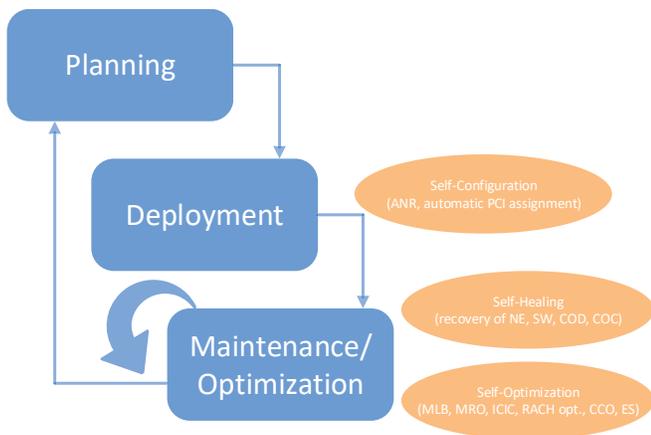}
\caption{Self-organizing networks}
\label{fig:sonn}
\end{figure}

SON has been introduced by \ac{3GPP} as a key component of \ac{LTE} network starting from the first release of this technology in Release 8, and expanding to subsequent releases. In SOCRATES \cite{SOCRATES} and in 3GPP \cite{3GPPUSES}, meaningful SON use cases have been defined, which can be classified according to the phases of the life cycle of a cellular systems (planning, deployment, maintenance and optimization) into: self-configuration, self-healing and self-optimization, as depicted in Figure~\ref{fig:sonn}. In this section, first we give an overview of the evolution of SON in \ac{3GPP}. We go through self-configuration, optimization and healing functionalities, introducing the use cases that have been defined for each one of them. We discuss about the self-coordination problem, to handle the potential conflicts that may exist between the parallel execution of multiple SON functions. We present the \ac{MDT} functionality. Finally, we focus on and end-to-end vision by extending SON principles to the core, and we discuss the role of virtualized and software defined networks in the context of 5G Network Management. Notice that here we do not focus on the academic literature, as it has already been reviewed in other interesting works \cite{survey}. We focus on the taxonomy defined by 3GPP, on the related roadmap, and we pay attention to the market penetration.

\begin{table*}[t!]
\centering \caption{Evolution of SON in 3GPP}
\label{Table:1}
\begin{adjustbox}{width=\linewidth}
\begin{tabular}{p{1cm}p{8cm}p{6cm}p{4cm}}
\hline
 \rule{0pt}{1\normalbaselineskip}
\textbf{\textcolor{myGreen}{Release}} & \textbf{\textcolor{myGreen}{WI}}&\textbf{ \textcolor{myGreen}{Feature}} & \textbf{\textcolor{myGreen}{TS or TR}}\\
\hline
 \rule{0pt}{1\normalbaselineskip}
\textbf{\textcolor{myGreen}{Rel.8}}&SA5-SON concepts and requirements& SON concepts and requirements  &  \cite{TS32.500} \\
\hline
 \rule{0pt}{1\normalbaselineskip}
\textbf{\textcolor{myGreen}{Rel.8}}&SA5-Self establishment of eNBs& Self configuration &  \cite{TS32.501,TS32.502,TS32.503,TS32.531,TS32.532,TS32.533}\\
\hline
 \rule{0pt}{1\normalbaselineskip}
\textbf{\textcolor{myGreen}{Rel.8}}&SA5-SON \ac{ANR} list management& ANR, PCI & \cite{TS32.761,TS32.762,TS32.763,TS32.765} \\
\hline
 \rule{0pt}{1\normalbaselineskip}
\textbf{\textcolor{myGreen}{Rel.9}}&SA5: Study of \ac{SON} related \ac{OAM} Interfaces for HeNBs&SON related OAM Interfaces for HeNBs  &  \cite{TS32.821} \\
\hline
 \rule{0pt}{1\normalbaselineskip}
\textbf{\textcolor{myGreen}{Rel.9}}&SA5: Study of self-healing of SON&Self-healing management&  \cite{TR32.823} \\
\hline
 \rule{0pt}{1\normalbaselineskip}
\textbf{\textcolor{myGreen}{Rel.9}}&SA5:SON OAM aspects: Automatic radio network configuration data preparation&Automatic radio network configuration data preparation &  \cite{TS32.501,TS32.502,TS32.503} \\
\hline
 \rule{0pt}{1\normalbaselineskip}
\textbf{\textcolor{myGreen}{Rel.9}}&SA5:SON OAM aspects self-organization management&Self-optimization (MRO, MLB, ICIC)&  \cite{TS32.425} \\
\hline
 \rule{0pt}{1\normalbaselineskip}
\textbf{\textcolor{myGreen}{Rel.9}}&RAN3: Self-organizing networks&CCO, MRO, MLB, RACH opt.&  \cite{TS25.413,TS36.300,TS36.413,TS36.423}\\
\hline
 \rule{0pt}{1\normalbaselineskip}
\textbf{\textcolor{myGreen}{Rel.10}}&SA5: SON self-optimization management continuation& Self-coordination, self-optimization (MRO, MLB, ICIC, RACH opt.)&  \cite{TS32.425,TS32.522,TS32.526,TS32.762,TS32.766} \\
\hline
 \rule{0pt}{1\normalbaselineskip}
\textbf{\textcolor{myGreen}{Rel.10}}&SA5: Self-healing management&CCO, COC&  \cite{TS32.541} \\
\hline
 \rule{0pt}{1\normalbaselineskip}
\textbf{\textcolor{myGreen}{Rel.10}}&SA5: OAM aspects of ES in radio networks&ES&  \cite{TR32.826,TR32.834,TS32.551,TS32.425,TS32.762,TS32.763,TS32.765} \\
\hline
 \rule{0pt}{1\normalbaselineskip}
\textbf{\textcolor{myGreen}{Rel.10}}&RAN2-3: LTE SON enhancements&CCO, ES, MLB, MRO enhancements&  \cite{TS36.300,TS36.331,TS36.413,TS36.423} \\
\hline
 \rule{0pt}{1\normalbaselineskip}
\textbf{\textcolor{myGreen}{Rel.11}}&SA5: ULTRAN SON management&SON management&  \cite{TS32.405,TS32.500,TR32.511,TS32.521,TS32.522,TS32.526,TS32.642}\\
\hline
 \rule{0pt}{1\normalbaselineskip}
\textbf{\textcolor{myGreen}{Rel.11}}&SA5: LTE SON coordination management&SON coordination \cite{S5-122330} &  \cite{TS32.425,TS32.500,TS32.521,TS32.522,TS32.526,TS32.762} \\
\hline
 \rule{0pt}{1\normalbaselineskip}
\textbf{\textcolor{myGreen}{Rel.11}}&SA5: Inter RAT ES management&OAM aspects of ES management&  \cite{TS32.405,TS32.551,TS32.522,TS32.526,TS32.642,TS32.646}\\
\hline
 \rule{0pt}{1\normalbaselineskip}
\textbf{\textcolor{myGreen}{Rel.11}}&RAN3: Further SON enhancements&MRO, MDT enhancements&  \cite{TS25.331,TS25.401,TS25.410,TS25.413,TS25.423,TS36.300,TS36.331,TS36.413,TS36.423}\\
\hline
 \rule{0pt}{1\normalbaselineskip}
\textbf{\textcolor{myGreen}{Rel.12}}&SA5: Enhanced NM centralized CCO&Enhanced NM centralized CCO&  \cite{TS32.836,TS32.425,TS32.103,TS28.627,TS28.628,TS28.658,TS28.659}\\
\hline
 \rule{0pt}{1\normalbaselineskip}
\textbf{\textcolor{myGreen}{Rel.12}}&SA5: Multi-vendor plug and play eNB connection to the network&Multi-vendor plug and play eNB connection to the network&  \cite{TS32.501,TS32.508,TS32.509}\\
\hline
 \rule{0pt}{1\normalbaselineskip}
\textbf{\textcolor{myGreen}{Rel.12}}&SA5: Enhancements on OAM aspects of distributed MLB&OAM aspects of distributed MLB & \cite{TR32.838}  \\
\hline
 \rule{0pt}{1\normalbaselineskip}
\textbf{\textcolor{myGreen}{Rel.12}}&SA5: Energy efficiency related performance measurements&Energy efficiency related performance measurements&  \cite{TS32.425} \\
\hline
 \rule{0pt}{1\normalbaselineskip}
\textbf{\textcolor{myGreen}{Rel.12}}&SA5: Het-Nets management$/$OAM aspects of network sharing&Het-Nets$/$network sharing &  \cite{TR32.835,TR32.851} \\
\hline
 \rule{0pt}{1\normalbaselineskip}
\textbf{\textcolor{myGreen}{Rel.12}}&RAN2-3: Next generation SON for ULTRAN$/$EUTRAN&SON per UE type, active antennas, small cells&  \cite{TR37.822} \\
\hline
 \rule{0pt}{1\normalbaselineskip}
\textbf{\textcolor{myGreen}{Rel.12}}&RAN2-3: ES enhancements for EUTRAN&ES&  \cite{TR36.887} \\
\hline
 \rule{0pt}{1\normalbaselineskip}
\textbf{\textcolor{myGreen}{Rel.13}}&RAN2-3: Enhanced Network Management centralized CCO&CCO& \cite{TS28.627}\\
\hline
 \rule{0pt}{1\normalbaselineskip}
\textbf{\textcolor{myGreen}{Rel.13}}&SA5: Study on Enhancements of OAM aspects of Distributed Mobility Load Balancing SON function& MLB&\cite{TR32.860}\\
\hline
 \rule{0pt}{1\normalbaselineskip}
\textbf{\textcolor{myGreen}{Rel.14}}&RAN: \ac{OAM} (SON for \ac{AAS}-based deployments) &Energy efficiency& \cite{rel14, ran14}\\
\hline
\end{tabular}
\end{adjustbox}
\end{table*}

\subsection{SON evolution in 3GPP}
\label{sec:SONevol}
\ac{3GPP} Release 8 started defining LTE and already sets the basis for concepts and requirements, and for SON functionalities regarding self-configuration, initial equipment installation and integration. The \ac{ANR} functionality is introduced here to reduce manual work when configuring the neighbouring list in newly deployed eNBs. Concepts of self-optimization are defined in the context of Release 9. It includes optimisation of coverage, capacity, handover and interference. The functions which are introduced (and that will be detailed in the following sections) are \ac{MLB}, \ac{MRO}, \ac{ICIC} and \ac{RACH} optimization. Release 10 focuses on enhancements to already defined SON functions to enhance interoperability between small cells and macro-cells and includes NGMNs recommendations, i.e., new functionalities such as \ac{CCO}, enhanced \ac{ICIC}, and it defines all the concepts related to self-healing, so \ac{COD} and \ac{COC} functions. Finally, concepts of \ac{MDT} and \ac{ES} are also introduced and then enhanced in Release 11. Release 11 SON functions are related to the automated management of heterogeneous networks. It includes mobility robustness optimization enhancements and
inter-radio access technology \ac{HO} optimization. Release 12 introduces optimization and enhancements for small cells including deployments in dense areas. In Release 13 novel concepts of unlicensed LTE have been introduced. Besides that, Release 13 studied the enhancements of \ac{OAM}, with respect to centralized and distributed architecture. In particular, focuses on distributed \ac{MLB}, as well as on enhanced NM or centralized CCO. Finally, Release 14 focuses on meeting the 5G requirements in terms of latency reduction, use of unlicensed spectrum in a fair manner, support for carrier aggregation, energy efficiency at OAM level, SON for active antennas, etc. Table~\ref{Table:1} summarizes the evolution of SON in 3GPP. Other documents of interest also include the protocol neutral SON policy \ac{NRM} \ac{IRP}, with the \ac{IS} \cite{TS32.522,TS32.762} and \ac{SS} \cite{TS32.526,TS32.766}.

\begin{figure}[t!]
\centering
\includegraphics[width=\linewidth]{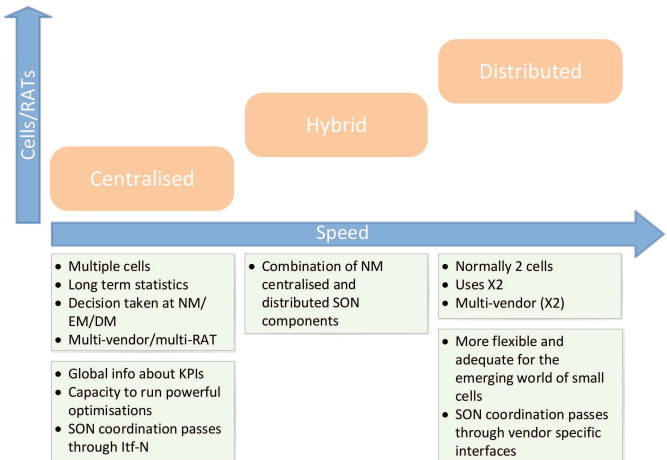}
\caption{SON implementations.}
\label{fig:son}
\end{figure}

\subsection{Self Configuration}
\label{sec:sc}
Self-configuration is the process of bringing a new network element into service with minimal human operator intervention \cite{TS32.501}. This covers the cellular system life cycle phase related to planning and deployment. Self-configuring algorithms take care of all configuration aspects of the \ac{eNB}. When the \ac{eNB} is powered on, it detects the transport link and establishes a connection with the core network elements. After this, the \ac{eNB} is ready to establish \ac{OAM}, S1 and X2 links and finally sets itself in operational mode. After the \ac{eNB} is configured, it performs  a self-test to deliver a status report to the network management node.
Since Release 8 \ac{ANR} and \ac{PCI} use cases have been considered \cite{TR30.818,anrpci}. The ANR function resides in the eNB and manages the conceptual Neighbour Relation Table (NRT). Located within ANR, the Neighbour Detection Function finds new neighbours and adds them to the NRT. ANR also contains the Neighbour Removal Function which removes outdated NRs. The Neighbour Detection Function and the Neighbour Removal Function are implementation specific \cite{parodi}. The \ac{PCI} is a physical layer signature to distinguish signals from different eNBs. It is based on synchronization signals. The total number of \ac{PCI}s is LTE is 504, so that reuse is inevitable, especially in dense deployments. The Automatic \ac{PCI} assignment aims at an automatic conflict and confusion free identification of cells \cite{TR36.902}.
Recommended practices for both use cases can be found in \cite{ngmn14}.
\begin{figure*} [t!]
\centering
\includegraphics[width=.8\textwidth]{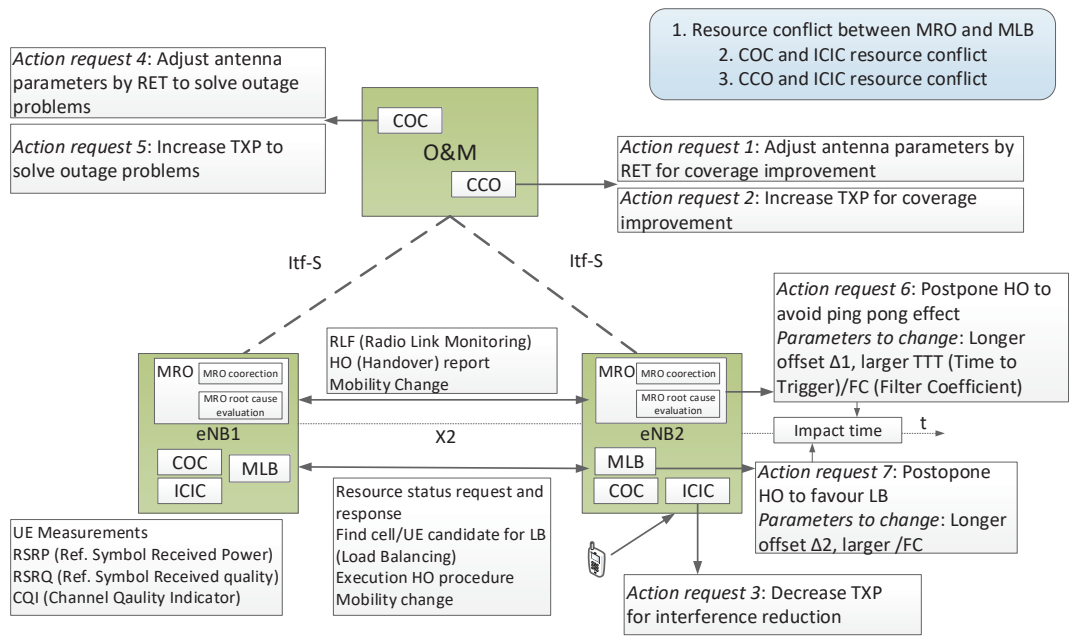}
\caption{High-level example of how the iterations of multiple SON functions may interfere.}
\label{fig:sonConflicts}
\end{figure*}

\subsection{Self Optimization}
\label{sec:sopt}
Self-optimization embraces all the set of mechanisms which optimize the network parameters during operation, based on measurements received from the network. In the following we provide a brief overview of the main self-optimization function that have been introduced across the different recent releases \cite{TR36.902}. From Release 9, we highlight work on:
\begin{enumerate}
\item \ac{MLB}. The \ac{MLB} is the \ac{SON} function in charge of managing cells' congestion through load transfer to other cells. The main objective is to improve the end-user experience and achieve higher system capacity by distributing user traffic across the system radio resources. The implementation of this function is generally distributed and supported by the load estimation and resource status exchange procedure. The messages containing useful information for this SON function (resource status request, response, failure and update) are transmitted over the X2 interface \cite{TS36.423}. \ac{MLB} can be implemented by tuning the \ac{CIO} parameter. The \ac{CIO} contains the offsets of the serving and the neighbour cells that all UEs in this cell must apply in order to satisfy the A3 handover condition \cite{TS36.213}.
\item \ac{MRO}. The \ac{MRO} is a SON function designed to guarantee proper mobility, i.e. proper handover in connected mode and cell re-selection in idle mode.  Among the specific goals of this function we have the minimization of call drops, the reduction of \acp{RLF}, the minimization of unnecessary hand-overs, ping-pongs, due to poor handover parameters settings, the minimization of idle problems. Its implementation is commonly distributed. The messages containing useful information are: the S1AP handover request or X2AP handover request, the handover report, the \ac{RLF} indication/report. Release 11 focused on different improvements of the handover optimization \cite{4gngmn}. \ac{MRO} operates over connected mode and idle mode parameters. In connected mode, it tunes meaningful handover trigger parameters, such as the event A3 offset (when referring to intra-RAT, intra-carrier hand-overs), the \ac{TTT}, or the Layer 1 and Layer 3 filter coefficients. In idle mode, it tunes the offset values, such as the Qoff-set for the intra-RAT, intra-carrier case.
\item \acf{ICIC}. \ac{ICIC} aims to minimize interference among cells using the same spectrum. It involves the coordination of physical resources between neighbouring cells to reduce interference from one cell to another. \ac{ICIC} can be done in both uplink and downlink for the data channels \ac{PDSCH}, and \ac{PUSCH}, or uplink control channel \ac{PDCCH}. \ac{ICIC} can be static, semi-static or dynamic. Dynamic \ac{ICIC} relies on frequent adjustments of parameters, supported by signalling among cells over X2 interface. To support proactive coordination among cells the \ac{HII} and the \ac{RNTP} indicators have been defined, while to support reactive coordination, the \ac{OI} has been introduced \cite{TS36.423}.
 \item \ac{RACH}. \ac{RACH} optimization aims at optimizing the random access channels in the cells based on UE feedback and knowledge of its neighbouring \acp{eNB} \ac{RACH} configuration. \ac{RACH} optimization can be done by adjusting the \ac{Pc} parameter or change the preamble format to reach the set target access delay \cite{36.300}.
\end{enumerate}
In Release 10, \ac{3GPP} defined new use cases.
\begin{enumerate}
\item \acf{CCO} is a SON function that aims to design self-optimizing algorithms that achieve optimal trade-offs between coverage and capacity. Different mechanisms can be considered to dynamically improve coverage and capacity, such as ICIC, scheduling, and the combination of such mechanisms. The targets that can be optimized may be vendor dependent and include coverage, cell throughout, edge cell throughput, or a weighted combination of the above.
\item \ac{ES} aims at providing the quality of experience to end users with minimal impact on the environment.
The objective is to optimize the energy consumption, by designing \acp{NE} with lower power consumption and temporarily shutting down unused capacity or nodes when not needed \cite{TR32.826}. In particular, many works in literature have been focusing on switching ON/OFF eNBs or small cells, in an efficient way, in order to guarantee a target level of Quality of service/experience, while minimizing the dissipated energy.
\end{enumerate}
Release 11 provides enhancements to MLB optimization, HO optimization, CCO, and ES. Release 12 has focused on a study on enhancements of OAM aspects for distributed MLB \cite{TR32.860}.
\subsection{Self-healing}
\label{sec:sh}
Self-healing \cite{TS32.541} focuses on the maintenance phase of a cellular network. Wireless cellular systems are prone to faults and failures, and the most critical domain for fault management is the \ac{RAN}. Every \ac{eNB} is responsible for serving an area, with little or none redundancy. If a \ac{NE} is not able to fulfill its responsibilities, it results in a period of degradation of performances, during which users are not receiving a proper service. This results in severe revenue loss for the operator.
Self-healing was initially studied in Release 9 \cite{TR32.823}, but it is in Release 10, when the main work has been carried out and features for detection, and adjustment of parameters have been specified \cite{S5-460036}. These specifications have been further updated in Release 11 \cite{TS32.541}. The main defined use cases are the following.
\begin{enumerate}
\item Self-recovery of NE Software. If the NE software failed due to load earlier software version and$/$or configuration, the most important thing is to ensure that the NE runs normally by removing the fault software, and restoring the configuration.
\item Self Healing of board Faults.  This use case aims to solve hardware failures in the NE \cite{son}.
\item Cell Outage Management. This use case is split in two main functions:
1) Cell Outage Detection. The main objective here is to detect a cell outage through the monitor performance indicators, which are compared against thresholds and profiles, and 2) Cell Outage Compensation. This use case aims at alleviating the outage caused by the loss of a cell from service \cite{TS32.541}. It refers to the automatic mitigation of the degradation effect of the outage by appropriately adjusting suitable radio parameters, such as the pilot power and the antenna parameters of the surrounding cells.
\end{enumerate}

\subsection{Self Coordination}
\label{sec:coord}
SON functionalities are often designed as stand-alone functionalities, by means of control loops. When they are executed concurrently in the same or different network elements, the impact of their interactions is not easy to be predicted, and unwanted effects may even occur among instances of the same SON function, when implemented in neighbouring cells. The risk of unacceptable oscillations of configuration parameters or undesirable performance results increase with the number of SON functions.

\ac{3GPP} has proposed different architectures for SON implementation, ranging from centralized C-SON to distributed D-SON. The choice of the architecture has a strong impact on the efficiency of the self-coordination framework. If C-SON is used, SON functions are implemented in the \ac{OMC} or in the \ac{NMS}, as part of the \ac{OSS}. This implementation benefits from global information about metrics and \ac{KPI}s, as well as computational capacity to run powerful optimization algorithms involving multiple variables or cells. However, it suffers from long time scales. In order to avoid oscillations of decision parameters, \ac{3GPP} requires \cite{S5-122330} that each SON function asks for permission before changing any configuration parameter. This means that a request must be sent from the \ac{SON} function to the SON coordinator and a response has to be returned. In \ac{C-SON} all these requests must pass through the Interface-N, which is not suitable for real-time communication, so that there is no possibility to give priority to \ac{SON} coordination messages over other \ac{OAM} messages. If in turn, distributed coordination is used, the interaction between the SON function and the local \ac{SON} coordinator will be over internal vendor-specific interfaces, with much lower latency characteristics. This makes the \ac{D-SON} architecture much more flexible and adequate for small cell networks, which experience very transitory traffic loads, thus requiring high reactivity to propagation and traffic conditions. 

An example of this can be observed in Figure~\ref{fig:sonConflicts}, where we provide an analysis of how the iterations among several SON functions implemented in centralized and distributed manner can generate conflicts in the network. In particular, this figure focuses on the SON output parameter conflict, i.e., when two or more SON functions aim at optimizing the same output parameter with different actions request, and where at least three possible conflicts can  arise: 1) the resource conflict between MRO and MLB; 2) the one among CCO and ICIC, and/or 3) the one among COC and ICIC use cases. We can identify output parameters, which are affected by two opposite decisions of two different functions, trying to achieve their own targets. As a result, to define and implement a self-coordination framework is considered a necessity \cite{schmelz,SOCRATES},\cite{altmanCoord}.

Market implementations of \ac{C-SON} are offered by vendors like Celcite (acquired by AMDOCS), Ingenia Telecom and Intucell (acquired by Cisco), while \ac{D-SON} solutions have traditionally been more challenging to implement and vendor specific, not allowing for easy interaction of products from different vendors, so that a supervisory layer is commonly still needed to coordinate the different instances of \ac{D-SON} across a much broader scope and scale. Only recently, vendors like Qualcomm or Airhop have started proposing \ac{D-SON} as a \ac{SON} mainstream, as small cells and \ac{Het-Net} require the millisecond response times of \ac{D-SON}.

\subsection{Minimization of Drive Tests}
\label{sec:mdt}
\ac{MDT} enables operators to collect \acp{UE} measurements together with location information, if available, with the purpose of optimizing network management while reducing operational effects and maintenance costs. This feature has been studied by \ac{3GPP} since Release 9 \cite{TR36.805}, among the targets there are the standardization of solutions for coverage optimization, mobility, capacity optimization, parametrization of common channels, and QoS verification \cite{son}. Since operators are also interested in estimating \ac{QoS} performance, in Release 11, MDT functionality has been enhanced through QoS performance to properly dimension and plan the network by collecting measurements indicating throughput and connectivity issues \cite{MDT3GPP}.
These MDT functions have been further elaborated in Release 11, while Release 12 has included specific enhancements in terms of correlation of information, which can be found in the study on enhanced network management centralized CCO. These improvements and extensions of SON enhancements introduced until Release 13 can be found in \cite{TS32.500}.

\subsection{Core networks}
\label{subsec:core}
The core network operations can be managed through self-organizing functionalities. The benefits also in this case come from the reduced human intervention and from reduced operational costs. self-organization in the core network allows to self-adapt traffic loads and prevent bottlenecks. In addition, self-organization for Core enables the core network to handle signalling more efficiently. In this regard, Nokia \cite{nokiacore} already automates core networks operations based on SON technology. The objective is to automatically and rapidly allocate core network resources to meet unpredictable behaviours and demands in terms of broadband. Notice that SON use cases for core networks are not limited to LTE networks, but many of them can be taken into account also for other kinds of networks, like 2/3G.

\subsection{Virtualized and Software defined networks}
\label{sec:nvf}
The wireless industry is currently working towards being prepared for a 1000x data traffic growth. It is unlikely, though, that users will want to pay more for the service than they are paying today, which set a serious challenge for both mobile operators and vendors, i.e. how to improve the infrastructure 1000 times, without increasing the CAPEX and OPEX. Besides SON, another trend in this direction, initiated by an ETSI industrial study group in 2012, is the \ac{NFV}, which allows to exploit the economies of scale of the IT industry, by moving traditional network functions away from specialized hardware to general purpose computation, storage and memory pools, distributed throughout the network and in data centers. NFV virtualizes the functional elements of the network, instantiating the corresponding functions as programs that run on commercial off-the shelf, and less expensive hardware. This concept, combined with a \ac{SDN} architecture, is introduced to make mobile network deployments more cost-effective \cite{SDN-survey},\cite{survey-NFV}.

The main idea behind these novel architectures is to provide a framework capable of assisting network operators to solve management problems, such as, cyber attacks, network failures, optimization to improve network performance, and \ac{QoE} of the users, among others. In this context, SON can be useful to achieve real time autonomous network management. In this novel softwerized visions, we can benefit of all the opportunities offered by centralized, distributed and local implementations proposed for SON at \ac{RAN} level, to extend this view beyond the radio access border, by proposing a SON over NFV architecture, where SON functions, aimed at tackling the main radio access and backhauling challenges of extremely dense deployments, are virtualized and run over generic purpose hardware. The NFV infrastructure is to be managed by an orchestrator entity, as proposed in ETSI architecture. Out of all the NFV architecture entities, this is the brain with the broadest view of the vertical service characteristics and the resource availability in the network. Therefore, it coordinates the  allocation of functions across the different segments of the dense, heterogeneous network. At the methodological level, the orchestrator can take advantage of the huge amount of information travelling through the network, in terms of measurements, signaling information, QoS and QoE indicators, etc., by means of machine learning based approaches.

In the market there already exist start-ups which advertise the concept of C-SON in the cloud. SON over NFV eliminates software and hardware dependencies, besides system scaling limitations, and offers reduction of costs through automatic processes. Cellwize \cite{cellwize} is one of them. They are promising a technology with deployment in the cloud, capable of working seamlessly across different vendors, spectrum and technologies. This research line is extremely novel and not much work can be found so far. However, we highlight the work that is under development in the context of the article and COGNET projects \cite{SELFNET}, \cite{COGNET}.

\section{How to address SON and NM through ML}
\label{sec:ml}
In this section we classify at high level the different network management classes of problems that one may need to deal with when aiming at managing the network in a self-organized manner. For each class of problem, we identify the machine learning tools that can be used.
The objective of ML is to improve performance of a particular sets of tasks by creating a model that helps find patterns through learning algorithms.
ML taxonomy is traditionally organized onto:
1. \ac{SL}, 2. \ac{UL}, 3. \ac{RL}. Recently, new trends in the area of ML are taking momentum, thanks to the progress of software engineering, computational capabilities and memory availability. Deep learning has been proven feasible and extremely effective in different applications, like language, video, speech recognition, object and audio detection, among others. The most exemplary one is the win of AlphaGo, beating the world champion at the Chinese board game Go. The victory of AlphaGo was due to the implementation of a deep reinforcement learning algorithm capable of self-learning.

Keeping in mind the SON and NM functions introduced in the previous section, the classes of problems that need to be addressed when managing the network autonomously are:

\begin{itemize}
\item \emph{Variable estimation or classification}: The tasks belonging to this class of problem aim at e.g. estimating the QoS or the QoE of the network, at predicting performances or behaviours of the network, by learning from the analysis of data obtained from past behaviours of the network. NM and SON functions where these tasks are useful are QoS estimation and other MDT use cases, the prediction of behaviours to optimize network parameters, etc. Solutions to these problems can be translated into finding the relationship between one variable and some others, or Identifying which class of a set of pre-defined classes the data belongs to. Solutions are then to be found in the SL literature, with both regression and classification tasks.
\item \emph{Diagnosis of network faults or misbehaviours}: The tasks belonging to this class of problems aim at detecting issues ongoing in the network, which may be associated to faults and anomalous setting of network parameters. This kind of problems relates to self-healing issues and solutions can be found in UL literature, and in particular in the anomaly detection solution.
\item \emph{Dimensionality reduction}: The network generates continuously a huge amount of data. For an appropriate processing and to extract useful information, it is convenient to eliminate the noise present in the data base, by reducing the dimensionality of data. Solutions to this problem are to be found in the UL literature, and specifically among the dimensionality reduction solutions.
\item \emph{Pattern identification, grouping}: The tasks belonging to this class aim at identifying patterns, group of nodes with similar characteristics, according to some kind of criteria. An objective may be to apply to them similar optimization approaches. Self-configuration use cases are intuitive application for these issues. Solutions to these problems can be translated into learning the set of classes the data belongs to.  UL literature offers solutions in the area of clustering.
\item \emph{Sequential decision problems for online parameter adjustment}: This class of problems is extremely common in the area of autonomous management, where we face control decision problems to online adjust network parameters, with the objective to meet certain performance targets. This kind of decision problems, where we learn the most appropriate decision online, based on the reaction of the environment to the actions the network is taking, can be addressed through RL solutions. All self-optimization use cases can be addressed through these solutions, as well as COC problems.
\end{itemize}
In the rest of the section, we relate each class of NM problem to the possible ML literature to solve it. The review of ML literature provided in the following, is far from being exhaustive. Many methods and techniques will not be described, because the purpose is here to provide a useful taxonomy to address NM and SON problems and to analyze and understand the related literature using ML solutions. For a deeper understanding of ML solutions, the reader is referred to more specific literature.

\subsection{\acf{SL}}
\label{subsec:prob2}
This ML technique could be extremely useful when the NM function to address requires estimation, prediction, classification of variables.
SL is a ML technique which takes training data (organized into an input vector ($\textbf{x}$) and a desired output value ($y$)) to develop a predictive model, by inferring a function $f(\textbf{x})$, returning the predicted output $\hat{y}$.
For that, the construction of a dataset is needed. The dataset contains training samples (rows), and features (columns), and is usually divided into 2 sets. The training set, used to train the model, and the test set, used to make sure that the predictions are correct. The goal of the training model is to minimize the error between the predictions and the actual values. Hence, by applying ML, we aim to estimate how well a learning algorithm generalizes beyond the samples in the training set.
The input space is represented by a $n$-dimensional input vector $\textbf{x}=(x^{(1)},\ldots, x^{(n)})^T \in \mathbf{R}^n$. Each dimension is an input variable. In addition a training set involves $m$ training samples $((\textbf{x}_1,y_1),\ldots,(\textbf{x}_m,y_m))$. Each sample consists of an input vector $\textbf{x}_i$, and a corresponding output $y_i$. Hence $x^{(j)}_i$ is the value of the input variable $x^{(j)}$ in training sample $i$, and the error is usually computed via $|\hat{y_{i}}-y_i|$. The SL technique has two main applications, classification and regression. On the one hand, classification is applied when $y$, the output value we try to predict is discrete, e.g., we want to predict if a cancer is benign or malign, based on a dataset constructed based on medical records, and collecting many features, e.g. tumour size, age, uniformity of cell size, uniformity of cell shape. On the other hand, a regression problem is applied when $y$ is a real number.

A huge amount of SL algorithms for classification can be found in the literature, and a study to evaluate the performance of some of them can be found in \cite{caruana}.
In the following we briefly introduce the most common algorithms.
\begin{enumerate}
\item \ac{$k$-NN} can be used for classification and regression. $k$-NN is a non-linear method where the input consists of the $k$ closest training samples in the input space. The predicted output is the average of the values of its $k$ nearest neighbours. A commonly used distance metric for continuous variables is the Euclidean distance. The $k$-NN method has the advantage of being easy to interpret, fast in training, and the amount of parameter tuning is minimal. However, the accuracy of the prediction is generally limited.
\item \ac{GLM}. The linear model describes a linear relationship between the output and one or more input variables, and where the approximation function maps from $x_i$ to $y_i$ as follows,
\begin{equation}
\hat{y_i}=\theta_0+\theta_1 x_i^{(1)}+\ldots +\theta_n x_i^{(n)}
\end{equation}
where $\theta_i$ are the unknown parameters. The idea is to choose $\theta_i$ so that $\hat{y_i}$ minimizes the loss function. Typically, we make the assumption that the samples in each dataset are independent from each other, and that the training set and testing set are identically distributed. Note that if the relation is not linear, the model should be generalized, in an attempt to capture this relationship \cite{glm}.
\item Naive Bayesian. The method is used for classification and is based on Bayes theorem, i.e., calculating probabilities based on the prior probability. The main task is to classify new data points as they arrive. A NB classifier assumes that all attributes are conditionally independent, and is recommended when the dimensionality of the input is high \cite{nb}. Since NB assumes independent variables, it only requires a small amount of training data to estimate the means and variances of the variables.
\item \ac{SVM} can be used for classification and regression. \ac{SVM} are inspired by statistical learning theory, which is a powerful tool for estimating multidimensional functions \cite{statistical,smola}. This method can be formulated as a mathematical optimization problem, which can be solved by known techniques. For this problem, given $m$ training samples $((\mathbf{x}_1,y_1),\ldots,(\mathbf{x}_m,y_m))$, the goal is to learn the parameters of a function which best fit the data. It samples hyperplanes. Thus, the hyperplane with the main minimum distance from the sample points is maintained. The sample points that form margin are called support vectors and establish the final model. This method in general shows high accuracy in the prediction, and it can also behave very well with non-linear problems when using appropriate kernel methods. Also, when we cannot find a good linear separator, kernel techniques are used to project data points into a higher dimensional space where they can become linearly separable. Hence the correct choice of kernel parameters is crucial for obtaining good results. In practice, this means that an exhaustive search must be conducted on the parameter space, thus complicating the task \cite{andreas}.

\item \ac{ANN} is a statistical learning model inspired by the structure of a human brain, where the interconnected nodes represent the neurons producing appropriate responses. ANN supports both classification and regression algorithms. The basic idea is to efficiently train and validate a neural network. Then, the trained network is used to make a prediction on the test set.
In this method the weights are the parameters in charge of manipulating the data in the calculations. Here, the interconnection pattern between the different layers of neurons, the learning process for updating the weights of the interconnections, and the activation function that converts a neuron's weighted input to its output activation are the most important parameters to be trained \cite{bishop}.
ANNs methods require parameters or distribution models derived from the data set, and in general they are also susceptible to over-fitting.

\item \ac{DT} is a flow-chart model in which each internal node represents a test on an attribute. Each leaf node represents a response, and the branch represents the outcome of the test \cite{idt}. DTs can be used for classification and regression, and they have nuisance parameters, such as the desired depth and number of leaves in the tree \cite{dt}. Also, they do not require any prior knowledge of the data, are robust (i.e., do not suffer the curse of dimensionality as they focus on the salient attributes) and work well on noisy data. However, DTs are dependent on the coverage of the training data as with many classifiers. Moreover, they are also susceptible to over-fitting.

\item \ac{HMM} can be used for classification, and also for other purposes. They can be used as a Bayesian classification framework, with a probabilistic model describing the data.

\end{enumerate}

Methodologies have also been proposed to take the best out of the available data, to boost the prediction performance. Some of these methodologies are classified among the so called \emph{Ensemble methods}. Ensemble methods combine the predictions of multiple learning algorithms to produce a final prediction. This technique has been investigated in a huge variety of works \cite{em,em2}. A general method is sub-sampling the training examples, where the most useful techniques are referred to as bagging and boosting \cite{em3}. Bagging manipulates the training examples to generate multiple hypothesis. It runs the learning algorithm several times, each one with different subset of training samples. On the other hand, AdaBoost maintains a set of weights over the original training set, and adjusts these weights by increasing the weight of samples that are misclassified, and decrease the weight of samples that are correctly classified \cite{mlresearch,freundSchapire}.

\subsection{\acf{UL}}
\label{subsec:prob3}

This kind of learning can be extremely useful when the NM function requires identifying anomalous behaviours, recognizing patterns or reducing the dimensionality of the data.
UL is a ML technique, which receives unlabelled input patterns with the objective to find a pattern in it. In this case, we let the computer learn by itself, without providing the correct answer to the problem we want to solve. The goal is to construct representation of inputs that can be used for predicting future inputs without giving the algorithm the right answer, as in turn we do in case of supervised learning \cite{ul}. The three most important families of algorithms are clustering, dimensionality reduction and anomaly detection techniques. There are many examples of UL applications in our daily life, e.g., news.google.com, understanding genomics, organize computer clusters, social network analysis, astronomical data analysis, market segmentation, etc. In the context of SON, UL algorithms are applied mainly on self-optimization and self-healing use cases.
\begin{enumerate}
\item Clustering. This technique aims at identifying groups of data to build representation of the input. The most common methods to create clusters by grouping the data are: non-overlapping, hierarchical and overlapping clustering methods. K-means \cite{kmeans} and \acp{SOM} \cite{som} methods belong to non-overlapping clustering techniques. When the clusters at one level are joined as clusters at the next level (cluster-tree), this is referred in literature as a hierarchical clustering method \cite{hc}. In case that an observation can exist in more than one cluster simultaneously, this is known as overlapping or fuzzy clustering. Fuzzy C-means and Gaussian mixture models belong to this kind of technique \cite{kmeans,fuzzyCmeans}. Also \ac{HMM} can be used for clustering This kind of algorithms have been proposed in a wide range of fields, such as, robotics, wireless systems, and routing algorithms for mobile ad-hoc networks, among others.
\item Dimensionality Reduction. High-dimensional datasets present many challenges. One of the problems is that, in many cases, not all the measured variables are necessary to understand the problem of interest. In the state of the art we can find a huge amount of algorithms to predict models with good performance from high-dimensional data. However it is of interest for many problems to reduce the dimension of the original data. For example, in \cite{jmoysenCAMAD, jmoysenHindawi}, the authors face the problem of the huge amount of potential features the system has as input, and they suggest that the regression analysis has a better performance in a reduced space. In this context, the most common methods are: \ac{FE} and \ac{FS} \cite{dr}. Both methods seek to reduce the number of features in the dataset. FE methods do so by creating new combinations of features (e.g. \ac{PCA}), which project the data onto a lower dimensional subspace by identifying correlated features in the data distribution. They retain the \acp{PC} with the greatest variance and discard all others to preserve maximum information and retain minimal redundancy \cite{pca}. Correlation based FS methods include and exclude features present in the data without changing them. For example, \ac{SPCA} extends the classic method of PCA for the reduction of dimensionality of data by adding sparsity constraint on the input features.

\item Anomaly Detection. Anomaly detection identifies events that do not correspond to an expected pattern. By modeling the most common behaviors, the machine selects the set of unusual events \cite{anomalyDet}. Self healing is one of the main functionality in which this kind of techniques are applied, some examples are \cite{banderaLG, munozLG}. The two most common techniques are:
\begin{itemize}
\item Rule based systems: they are very similar to DTs, but they are more flexible than DTs as new rules may be added, without creating a conflict with the existing ones \cite{anomalyDet}.
\item Pruning techniques: they aim at identifying outliers, where there are errors in any combination of variables.
\end{itemize}

\item Latent Variable models. This kind of techniques allows learning a model where some unseen variable helps simplify and describe the data. An example is the non negative matrix factorization.

\end{enumerate}

\subsection{\acf{RL}}
\label{subsec:prob1}
The ML approaches under this category can be used to address NM functions which require network parameter control.
Differently from the case of SL, \ac{RL} aims to learn from interactions how to achieve a certain goal. In many real applications and in particular, in sequential decision and control problems, it is not possible to provide an explicit supervision to the training (i.e. the right answer to the problem). In these cases, we can only provide a reward/cost function, which indicates to the algorithm when it is doing well and when it is doing poorly. RL has already been proven effective in many real world applications, such as autonomous helicopters, network routing, robot legged automation, etc. \cite{schaal, thrun, littman}.

The learner or decision maker is called \textit{agent}, and it interacts continuously with the so-called \textit{environment}. The agent selects actions and the environment responds to those actions and evolves into new situations. In particular, the environment responds to the actions through \textit{rewards}, i.e., numerical values that the agent tries to maximize over time.

The agent has to exploit what it already knows in order to obtain a positive reward, but it also has to explore in order to take better actions in the future. Learning can be centralized in a single agent or distributed across a multiple agents. In single agent systems, ML approaches are capable of finding optimal decision policies in dynamic scenarios with only one decision maker. In multi agent systems, the distributed decisions are made by multiple intelligent decision makers, and the optimal solutions or equilibria are not always guaranteed \cite{panait}.

The problem is then defined by means of a \ac{MDP} $\left\{ \mathcal{S}, \mathcal{A}, \mathcal{T}, \mathcal{R}, \gamma \right\}$, where $\mathcal{S}$, is the set of possible states of the environment $\mathcal{S}=\left\{s_1,s_2,\ldots,s_n\right\}$, $\mathcal{A}$,
is the set of possible actions $\mathcal{A}=\{a_1,a_2,\ldots,a_q\}$ that each decision maker may choose, $\mathcal{T}(s'|s,a)$, is the transition function denoting the probability of getting $s'$ giving an action $a$ in state $s$, $\mathcal{R}(s,a)$ is a reward function, which specifies the expected immediate return obtained by executing action $a$ in state $s$, and $0\leq\gamma\leq1$ is a discount factor, which gives more importance to immediate rewards compared to rewards obtained in the future \cite{suton}.

The \ac{MDP} represents the theoretical basis for the \ac{RL} framework \cite{suton}.
At each time step, the agent implements a mapping from states to probabilities of selecting each possible action. This mapping is the agent's \emph{policy}.

The objective of each learning process is to find an optimal policy $\pi^*(s) \in \mathcal{A}$ for each $s$, to maximize some cumulative measure of the reward $r$ received over time. Almost all \ac{RL} algorithms are based on estimating a so called \emph{value function}, which is a function of the states estimating how good it is for an agent to be in a given state. For MDPs the state-value function, denoted as $V_{\pi}(s)$, is the expected return when starting in state $s$ and following policy $\pi$ thereafter. For more information the reader is refereed to \cite{suton}.

\ac{RL} literature offers two approaches to solve \ac{MDP}s. These two approaches are: model-based and model-free.

\begin{enumerate}
\item Model-based. \ac{DP} and \ac{MC} methods fall into the category of model-based approach.
\begin{enumerate}
\item \ac{DP} is able to solve \ac{MDP}s by relying on the knowledge of the state transition probability between two states after executing a certain action. DP is an algorithmic paradigm that solves a given complex problem by breaking it into sub-problems and stores the results of sub-problems to avoid computing the same results again. DP algorithms are based on update rules derived from the Bellman equation. The first key component is known as the \emph{policy evaluation} process, according to which a policy provides information about how much reward is going to be received in the MDP. This solution is used to build the first overall solution by finding the optimal policy known as the \emph{policy iteration} process. Finally the \emph{value iteration} makes the value function better and better by applying Bellman's equation intuitively. DP is used to solve problems such as, scheduling, graph algorithms, bioinformatics, among others.
\item MC method only requires experience, i.e., sample sequences of states, actions and rewards. The estimations are only updated after the episodes conclude. Although their application on practical cases is limited, they provide foundation for other RL methods.
\end{enumerate}
\item Model-free. \ac{TD} methods are model free approaches to solve RL problems. TD learning is a combination of MC and DP ideas. It uses the current estimate $V^{\pi}_t$ of the value function instead of the exact $V^{\pi}$, as it happens in DP. If $\mathcal{T}$ is known, we can solve the MDP through DP, otherwise we need to rely on TD methods.

Some common examples of TD methods are: Q-learning, Sarsa and \ac{AC} \cite{suton}. TD methods can be found in each SON functionality.
\begin{enumerate}
\item Q-learning and Sarsa are based on the estimation of the state-action value function, $Q(s, a)$. Learning is performed by iteratively updating the Q-values, which represent the expert knowledge of the agent, and have to be stored in a representation mechanism. The most intuitive and common representation mechanism is the lookup table, i.e., the TD methods represent their Q-values in a Q-table, whose dimension depends on the size of the state and action sets. The difference between them is that, Q-learning is an off-policy learner. This means that, the agent will use the policy corresponding to the best action in the next state, given the current agent experience, whereas Sarsa is an on-policy learner. On-policy learners evaluate the policy $\pi$, to perform the decisions. This means that, the policy followed by the agent to select its behaviour in a given state is the same used to select the action based on which it evaluates the followed behaviour.
\item \ac{AC} methods have a separate memory structure to represent the policy independently of the value function. The policy structure is known as the \emph{actor}, since it is used to select the actions, while the estimated value function is known as the \emph{critic}.
The critic learns and critiques whatever policy is currently being followed by the actor and takes the form of a \ac{TD} error $\delta$, which is used to determine if $a_{t}$ was a good action or not. $\delta$ is a scalar signal, which is the output of the critic and drives the learning procedure. After each action selection, the critic evaluates the new state to determine whether things have gone better or worse than expected.
\end{enumerate}
\end{enumerate}

\section{Machine learning enabled Network Management}
\label{sec:MLinSON}

As we have mentioned in the introduction of this work, mobile networks constitute a huge source of data which could be analyzed with proper tools, with the primary goal to make more informed decisions when it comes to efficiently manage the overall 4G or 5G network. In this context, ML is a great opportunity due to its capability of providing insightful information from the analysis of data already available to operators, which can be used to make improvements or changes.

In this section we focus on how ML can specifically be applied to SON and novel network management concepts. First, we present all the relevant sources of information that could be extracted from a mobile network. All these data are available to operators, and may happen to be sensitive data for the users' privacy. However, some interesting data can be derived from open databases or sniffed from unencrypted control channels like the \ac{PDCCH}. We will then discuss on these options. Third, we will go through again the main SON and network management functions and provide a classification of the main inputs and outputs that we would need available in the form of data, when designing an appropriate ML algorithm to target the specific use case, and the \ac{KPI} indicators that we would need to monitor. Finally, we provide an overview of SON and network management's related work, where ML techniques have been adopted, classifying this work as a function of the targeted use case, the specific high level problem to solve and the ML technique that the authors have picked to address the problem.

\subsection{Data generated by mobile cellular networks}
\label{sec:sources}
As we observed in \cite{bDemp} a huge amount of data is currently already generated in mobile networks during normal operations by control and management functions. This kind of data can be exploited to find patterns and extract useful information from them. This allows to take more informed decisions to effectively manage network performance. Some examples of the different sources information generated by mobile networks, together with the kind of usage currently provided by operators, and related references of interest, is detailed in Table~\ref{Table:data}.
\begin{enumerate}
\item \acf{CDR}. They are defined in ~\cite{TS32.298} and provide a comprehensive set of statistics at the service, bearer and IP Multimedia System (IMS) levels. These records are typically stored for offline processing by the operator. The granularity of this information in the time domain is however quite coarse, as records are generated in correspondence with high-level service events (e.g., start of a call).
\item Performance management functionality. This data source ~\cite{TS32.401}~\cite{TS32.425} provides data regarding the network performance and it covers, among others, aspects of the performance of the radio access network, such as, radio resource control and utilization, performance of the various bearers (both on the radio part and in the back-haul), idle and connected mode mobility.
\item \acf{MDT}. The data extracted from this source refers to the radio measurements of both idle and connected mode mobility, coverage items, such as, power measurements and radio link failure events, and can be associated with position information of the UE performing the measurement. More information on these data has already been provided also in section \ref{sec:mdt}.
\item E-UTRA Control plane protocols and interfaces, such as \ac{RRC}, S1-AP, X2-AP protocols, are another huge source of information, especially concerning aspects, such as cell coverage, user connectivity, mobility in idle and connected mode, inter-cell  interference, resource management, load balancing, among others.
\item Data plane traffic flow statistics, also are a huge source of information, which can be gathered at various points of the network, like the eNB, or the \ac{PGW} and \ac{SGW}. The Internet Protocol Flow Information Export (IPFIX) is an example of standardized format to exchange this kind of statistics \cite{IETFRFC7012}.
\end{enumerate}

\begin{table*}[t!]
\centering
\caption{Information elements relevant for ML enabled SONs}
\label{Table:data}
\begin{adjustbox}{width=\linewidth}
\begin{tabular}{p{3cm}p{8cm}p{5.5cm}p{2cm}}
\arrayrulecolor{green}\hline
\textbf{\textcolor{myGreen}{Source}} & \textbf{\textcolor{myGreen}{Data}}&
\textbf{\textcolor{myGreen}{Usage}}
 &\textbf{\textcolor{myGreen}{TS}} \\
\hline  \rule{0pt}{1\normalbaselineskip}\acf{CDR} &
\begin{minipage}[t]{\linewidth}%
Includes statistics at the service, bearer and \ac{IMS} levels.
\end{minipage} & These records are typically
stored, but only used by customer service. The network operation departments typically do not leverage this information and do not have access to it, as much as customer service does not leverage network management data.
 & TS~32.298~\cite{TS32.298}
\\
\arrayrulecolor{black}\hline
 \rule{0pt}{1\normalbaselineskip}Performance management (data on network performance)&
\begin{minipage}[t]{\linewidth}%
It covers long-term network operation functionalities, such as Fault, Configuration, Accounting, Performance and Security management (FCAPS), as well as customer and terminal management. An example is that defined for Operations, Administration, and Management (OAM), which consists of aggregated statistics on network performance, such as number of active users, active bearers, successful/failed handover events, etc. per BS, as well as information gathered by means of active probing.
\end{minipage} & The data is currently mostly used for fault identification, e.g., triggering alarms when some performance indicator passes some threshold, so that an engineer can investigate and fix the problem. Typically, the only automatic use of this info is threshold-based triggering, which can be done with very low computational complexity. & TS32.401~\cite{TS32.401}, TS32.425~\cite{TS32.425}
\\
\arrayrulecolor{black}\hline
 \rule{0pt}{1\normalbaselineskip}\acf{MDT}  &
\begin{minipage}[t]{\linewidth}%
Radio measurements for coverage, capacity, mobility optimization, QoS optimization/verification
\end{minipage} &This data is used for identified use cases such as coverage, mobility and capacity optimization, and QoS verification & TS37.320~\cite{TS37.320}
\\
\arrayrulecolor{black}\hline
 \rule{0pt}{1\normalbaselineskip}E-UTRA Control plane protocols and interfaces &
\begin{minipage}[t]{\linewidth}%


Control information related to regular short-term network operation, covering functionalities such as call/session set-up, release and maintenance, security, QoS, idle and connected mode mobility, and radio resource control.

\end{minipage} &  A
This information is normally discarded after network operation purposes have been fulfilled. Some data can be gathered via tracing functionality or used by SON algorithms which normally discards the information after usage & TS36.331~\cite{TS36.331}, TS36.413~\cite{TS36.413}, TS36.423~\cite{TS36.423}
\\
\hline
\end{tabular}
\end{adjustbox}
\end{table*}

All these data are available to the network operators, but in most cases this is not made available to the academic community due to privacy issues and network operators' interests. There are some exceptions, like the \emph{Data for Development} (D4D) initiative from the Orange group~\cite{D4D}, which made available anonymous data extracted from the Senegal's network to research laboratories. However these data are in general aggregated and do not allow deep insight into the operator's network.

This lack of data represents a great limitation for the advancement of the ML based network management research. However, some network data can be derived in other ways.
Some databases are available, providing a huge insight in mobile network operators. Some examples are listed in the following, together with information that can be extracted from them.

\begin{itemize}
{
\item \textbf{opencellid}: It contains information on specific cells, such as: network type (GSM, UMTS, LTE), Mobile Country Code (MCC), Mobile Network Code (MNC), Location Area Code (LAC) for GSM and UMTS, Tracking Area Code (TAC) for LTE, Cell ID for (CID) for GSM and LTE networks, Primary Scrambling Code (PSC) for UMTS networks, Physical Cell ID (PCI) for LTE networks, longitude and latitude in degrees, estimates of range in meters, total number of measurements collected from the tower, defines if the coordinates of the cell tower are exact or estimated, information of the date when the cell tower was first added to the data base and updated, average signal strength from all measurements received from the cell in dBm, or as defined in \cite{TS27.007}. This data base also receives funding from important vendors like Qualcomm \cite{opencellID} and offers some formula of free access to portion of data for academic purposes.
\item \textbf{opensignal}: It offers information on achievable data rates, latency and availability, per operators, but not information per cell tower~\cite{opensignal}.
\item \textbf{antenasgsm}: It offers information on maps and positions of cells, with added information on the operator and the assigned bandwidth~\cite{antenasgsm}.
\item \textbf{Google geolocations API}: It allows queries based on the cell ID to get cell related information and WiFi Access Points (AP), such as latitude and longitude~\cite{googlegeo}.
}
\end{itemize}

The information provided by these databases is precious, but still does not give sufficient insight on the behaviour of the network, and mainly offers an overview of the coverage provided by the single operators. To get more information, still we can do something more and access directly to the unencrypted \ac{PDCCH} and extract information exchanged between the users and the associated eNB. In particular, it is possible to build a sniffer, as the one described in~\cite{BuiOWL}, from which to collect raw communication traces exchanged by the users and the
associated eNodeB. This allows to have access not only to aggregate base station statistics, but also to more valuable information derived from the radio protocols, such as the resource block allocation and the link adaptation mechanism of the system. In particular, the OWL sniffer~\cite{BuiOWL} is an online decoder of the LTE control channel, which uses a Software Defined Radio (SDR) to send the raw LTE signal to a PC running the decoding software. This open-source software is capable of reliably logging the LTE downlink control information (DCI) broadcasted by base stations. In
fact, LTE uses an unencrypted control channel to assign network resources to users for both downlink and
uplink communications. Resources are assigned to devices through their radio network temporary identifiers
(RNTIs), every millisecond, specifying the number of resource blocks (RBs) and the modulation and coding
scheme (MCS) to be used. There are works in literature using this sniffer to collect and analyze traces from different European cities~\cite{Trinh17}.

Finally, let us review the main SON use cases in Table \ref{Table:SON_IO_KPI}, by analyzing the main input information that their design would require, in terms of data, together with the main identified output actions and meaningful associated \ac{KPI}s.

\begin{table*}[t!]
\centering
\caption{SON inputs, outputs and KPIs}
\label{Table:SON_IO_KPI}
\begin{adjustbox}{width=\linewidth}
\begin{tabular}{p{4cm}p{5cm}p{6cm}p{4cm}}
\arrayrulecolor{green}\hline
\textbf{\textcolor{myGreen}{SON function}} & \textbf{\textcolor{myGreen}{Inputs}}&
\textbf{\textcolor{myGreen}{Output actions}}
 &\textbf{\textcolor{myGreen}{KPIs}} \\
\hline
 \rule{0pt}{1\normalbaselineskip}\acf{MLB} &
\begin{minipage}[t]{\linewidth}%
X2 resource status and load estimation information.
\end{minipage} & Tuning the \ac{CIO}, i.e. offsets of serving and neighbour cells to satisfy handover conditions.
 & Improved QoS and capacity
\\
\arrayrulecolor{black}\hline
 \rule{0pt}{1\normalbaselineskip}\acf{MRO}&
\begin{minipage}[t]{\linewidth}%
S1AP and X2AP handover requests, handover reports, \ac{RLF} reports and indications.
\end{minipage} & A3 offsets, \ac{TTT}, L1 and L3 filter coefficients, in connected mode, and Qoffset in Idle mode. & Minimized call drops, \ac{RLF}s and ping pong effects.
\\
\arrayrulecolor{black}\hline
 \rule{0pt}{1\normalbaselineskip}\acf{CCO}  &
\begin{minipage}[t]{\linewidth}%
UE measurements
\end{minipage} & Transmission power, pilot power, antenna parameters, coordinated \ac{ABS} & Maximized coverage and cell and edge throughput
\\
\arrayrulecolor{black}\hline
 \rule{0pt}{1\normalbaselineskip}\acf{ICIC} &
\begin{minipage}[t]{\linewidth}%
\ac{HII}, \ac{RNTP}, \ac{OI}, UE Measurements.
\end{minipage} & Transmission power, pilot power, antenna parameters, coordinated \ac{ABS} & Minimized Intercell interference.
\\
\arrayrulecolor{black}\hline
 \rule{0pt}{1\normalbaselineskip}\acf{COC} &
\begin{minipage}[t]{\linewidth}%
UE Measurements.
\end{minipage} & Transmission power, antenna parameters of neighbouring cells & Minimized outage.
\\
\arrayrulecolor{black}\hline
 \rule{0pt}{1\normalbaselineskip}\acf{ES} &
\begin{minipage}[t]{\linewidth}%
Resource status, UE Measurements.
\end{minipage} & Switch ON and OFF policies & Minimized energy consumption.
\\
\hline
\end{tabular}
\end{adjustbox}
\end{table*}

\subsection{Overview of ML based Network management's relevant literature}

This section reviews SON and Network management's recent work in the area of ML. We will go through each main function and use case and review significant literature and ML approach that has been used to approach the problem. Table~\ref{Table:relwork} summarizes the main works in this area and classifies them per \ac{3GPP} use case, technique and specific algorithm adopted by the authors.

\begin{enumerate}
\item Use case: Indicates the 3GPP targeted use case.
\item Reference: Indicates the reference of the related work.
\item Technique: Indicates the applied ML method (Supervised Learning, Unsupervised Learning, Reinforcement Learning).
\item Problem: Indicates the general problem to solve.
\item Algorithms: Indicates the specific ML algorithm applied to the data (see Table\ref{Table:relwork}).
\end{enumerate}

\begin{table*}[t!]
\centering
\caption{Related work}
\label{Table:relwork}
\begin{adjustbox}{width=\textwidth}
\begin{tabular}{lllll}
\arrayrulecolor{black}\hline
\rule{0pt}{1\normalbaselineskip}
& \textbf{\textcolor{myGreen}{Reference}} & \textbf{\textcolor{myGreen}{ML technique}} & \textbf{\textcolor{myGreen}{Problem}}& \textbf{\textcolor{myGreen}{Algorithm}}\\
\hline\hline
 \rule{0pt}{1\normalbaselineskip}\textbf{\textcolor{myGreen}{Self-configuration}} &&&&\\

PCI &\cite{Peng13} &UL & Planning  &  Clustering \\
\hline\hline
 \rule{0pt}{1\normalbaselineskip}\textbf{\textcolor{myGreen}{Self-optimization}} &&&&\\

 \rule{0pt}{1\normalbaselineskip}MLB &\cite{stephenMLB}&RL& Control optimization  &Q-learning\\
&\cite{munoz2}&RL& Control optimization       & Q-learning\\
&\cite{mlb} & RL & Control optimization  & Fuzzy Q-learning \\
&\cite{junishi}&RL&  Control optimization    &Dynamic Programming\\
&\cite{emil}&UL& Grouping  & K-means clustering\\
&\cite{Franco15}&SL& Prediction  & Multivariate polynomial regression \\
\hline
\rule{0pt}{1\normalbaselineskip}MRO 
&\cite{qin}&RL &    Control optimization   &Q-learning\\
&\cite{stephenMLB2}&RL&  Control optimization  &Q-learning\\
&\cite{HOpablo}&RL&  Control optimization  &Fuzzy control\\
&\cite{mro,mrosinclair}&UL& Pattern identification &SOM\\
&\cite{Farooq17}&UL&  Prediction  &Semi-Markov model\\

&\cite{Ali16, Ostlin04, Quintero04, Majumdar05}&SL&Prediction&ANN\\
\hline
 \rule{0pt}{1\normalbaselineskip}CCO & \cite{RAZAVI}&RL&  Control optimization    &Fuzzy Q-learning\\
& \cite{naseer}&RL&  Control optimization    &Fuzzy Q-learning\\
& \cite{jingyu} &   RL&   Control optimization     & Fuzzy Q-Learning \\
&\cite{ccoEURASIP}&UL, RL&  Control optimization  &Fuzzy ANN/Q-learning\\
\hline
 \rule{0pt}{1\normalbaselineskip}ICIC&\cite{Galindo}&RL&  Control optimization   &Q-learning\\
&\cite{dirani}&RL&     Control optimization      &Fuzzy Q-learning\\
&\cite{blascoICIC, simsek}&RL&  Control optimization   &Q-learning\\
\hline
 \rule{0pt}{1\normalbaselineskip}ES&\cite{miozzo}& RL&  Control optimization       & Q-learning\\
&\cite{annaCAMAD}& UL&  Decision making      & Fuzzy logic\\
&\cite{es1, es2}& UL&  Grouping, pattern identification     & Clustering\\

\hline\hline
\textbf{\textcolor{myGreen}{Self-healing}}\\

 \rule{0pt}{1\normalbaselineskip}COC
&\cite{jmoysenSH}&RL&  Control optimization    &Actor Critic\\
&\cite{onireti}&RL& Control optimization & Actor-Critic\\
&\cite{ARSALAN}&SL&    Control optimization   &Fuzzy logic\\
\hline

\rule{0pt}{1\normalbaselineskip}COD&\cite{fedor}&UL&  Anomaly detection  &Diffusion Maps\\
&\cite{khabib}&SL& Anomaly detection  &Fuzzy logic \\
& \cite{RANA}&SL/UL&Diagnosis &Naive Bayesian \\
&\cite{emcoc}&SL&  Anomaly detection &SVM, Ensemble methods\\
&\cite{onireti}&SL/UL&  Anomaly detection &$k$-NN, local-outlier-factor\\
&\cite{Alias16}&UL& Grouping, pattern identification &Hidden Markov Model\\
&\cite{Xue14,Zoha14,Chernov14}&SL&Fault Detection&$k$-NN\\
&\cite{Barco05,Khanafer08}&SL&Diagnosis&Naive Bayesian\\
\hline\hline
 \rule{0pt}{1\normalbaselineskip}\textbf{\textcolor{myGreen}{Self-coordination}}\\

&\cite{hafiz}&SL& Classification & Decision Trees\\
&\cite{Berna_coord}  & RL&   Control optimization  &Actor Critic\\
&\cite{lbHo}  &RL & Control optimization  &Q-learning\\
&\cite{jmoysenEurasip}  & RL& Control optimization  &Actor Critic\\
\hline\hline
 \rule{0pt}{1\normalbaselineskip}\textbf{\textcolor{myGreen}{Minimization Drive Tests}}\\

&\cite{chernogorov12,chernogorov13} &SL&   Verification/estimation &Linear correlation\\
&\cite{jmoysenCAMAD}&SL&   Prediction      &Regression models\\
&\cite{jmoysenISCC}&SL/UL&   Prediction/curse of dimensionality   &Regression models/Dimensionality reduction\\
&\cite{jmoysenPIMRC, jmoysenHindawi}&SL&   Prediction     &Bagged-SVM/Dimensionality reduction\\
\hline\hline
 \rule{0pt}{1\normalbaselineskip}\textbf{\textcolor{myGreen}{Core Networks}}\\

&\cite{balint} &SL&   Prediction     &Adaboost, SVM\\

\hline
\end{tabular}
\end{adjustbox}

\end{table*}

\subsubsection{Mobility Load Balancing}
The literature offers some examples of application of ML techniques to the MLB use case. The majority of applications fall in the area of RL, as the main problem to solve is a sequential decision problem about how to set configuration parameters, which optimize network performance and user experience.
An example of a RL application for MLB use case can be found in \cite{stephenMLB}. Here the authors present a distributed Q-learning approach that learns for each load state the best MLB action to take, while also minimizing the degradation in HO metrics. Another option to take advantage also of fuzzy logic capabilities of dealing with heterogeneous sources of information is provided in \cite{munoz2}, where fuzzy logic is combined with Q-learning in order to target the load balancing problem. For similar reasons, fuzzy logic is also proposed in \cite{mlb} to enhance the network performance by tuning HO parameters at the adjacent cells. Approaches incorporating fuzzy logic with RL  capabilities have the advantage to capture the uncertainty existing in real world complex scenarios, while schemes considering only learning approaches may be limited by the fixed variable definition. When combining fuzzy logic with RL, also the subjectivity with which the fuzzy variable may be defined is overcome by the adjusting capabilities of the learning. Alternatively, a centralized solution is approached in \cite{junishi}, where a central server in the cellular network determines all HO margins among cells by means of a dynamic programming approach. Besides RL, also clustering schemes have been proposed in this area, to group cells with similar characteristics and provide for them similar configuration parameters \cite{emil}. Considering clustering in large realistic scenarios is an added value to reduce computational complexity and take advantage of what is learnt in other regions of the network where we observe similar environment characteristics.

\subsubsection{Mobility Robustness Optimization}
Also for the case of MRO, we find in literature different solutions based on RL to solve a control decision problem.
In \cite{qin,stephenMLB2}, the authors focus on the optimization of the users' experience and of the HO performance.
In \cite{qin} the authors take advantage of the Q-learning approach to effectively reduce the call drop rates, whereas in \cite{stephenMLB2}, unlike other solutions that assume a general constant mobility, the authors adjust the HO settings in response to the mobility changes in the network by means of a distributive cooperative Q-learning. Differently from \cite{qin,stephenMLB2}, in \cite{HOpablo}, the authors take advantage also of fuzzy logic capabilities. These solutions are based on control optimization of HO parameters through RL, so they propose similar solutions to those found in the literature of MLB. In this case we can do the same considerations about the advantages of considering fuzzy logic in order to gain in flexibility in the uncertain and complex real network context. Different approaches in turn, address the problem by identifying successful HO events, through solutions based on unsupervised learning. In particular, the works of \cite{mro} and \cite{mrosinclair} propose an approach to HO management based on \ac{UL} and \ac{SOM} analysis. The idea is to exploit the experience gained from the analysis of data of the network based on the angle of arrival and the received signal strength of the user, to learn specific locations where HOs have occurred and decide whether to allow or forbid certain handovers to enhance the network performance. The solutions enable self-tuning of HO parameters to learn optimal parameters' adaptation policies. Similarly, in \cite{Farooq17} the authors exploit the huge amount of information generated in the network to predict user traffic distribution. In particular, they take advantage of semi-Markov model for spatiotemporal mobility prediction in cellular networks. Finally, the works in \cite{Ali16, Ostlin04, Quintero04, Majumdar05}, propose schemes to make predictions about UE's mobility, which allows to anticipate smart HO decisions.

\subsubsection{Coverage and Capacity Optimization}

In case of CCO, different approaches in literature focus on RL solutions based on continuous interactions with the environment, oriented to online adjusting antenna tilts and transmission power levels through TD learning approaches. In \cite{RAZAVI} and \cite{naseer} a fuzzy Q-learning approach to optimize the complex wireless network, by learning the optimal antenna tilt control policy has been proposed, and a similar approach is followed also in \cite{jingyu} and \cite{ccoEURASIP}. In addition, they also propose to combine fuzzy logic with Q-learning, in order to deal with continuous input and output variables. \cite{jingyu} also proposes a central control mechanism, which is responsible to initiate and terminate the learning optimization process of every learning agent deployed in each eNB.
Finally, \cite{ccoEURASIP} innovates with respect to other approaches since in order to adjust the antenna tilt and transmission power parameters, it considers the load distribution of the different cells involved in the optimization process, and introduces novel mechanisms to facilitate cooperative learning among the different SON entities.

\subsubsection{Inter-cell Interference Coordination}
Similarly to the CCO case, ML has been proposed in the literature of ICIC use case as a valid solution, where RL is the principle used tool, with special emphasis to TD methods, in order to target the optimization of control parameters.
Several works target the problem to minimize the interference among cells by using the most common TD learning method, Q-learning \cite{Galindo,dirani,blascoICIC,simsek}.
The work in \cite{Galindo} is related to control inter-cell interference in a heterogeneous femto-macro network. The work combines information handled by the multi-user scheduling with decisions taken by a learning agent based on Q-learning, which tries to control the cross-tier interference per resource block.
\cite{dirani} proposes a distributed solution for ICIC in OFDMA networks based on a Fuzzy Q-learning implementation. The proposed solution achieves joint improvement for all users, i.e., the improvements of users with bad quality does not come at the expense of users with good quality.
Moreover, a decentralised Q-learning framework for interference management in small cells is proposed in \cite{blascoICIC}. The authors focus on a use case in which the small cell networks aim to mitigate the interference caused to the macro-cell network, while maximizing their own spectral efficiencies. Finally,
in \cite{simsek} also a decentralized Q-learning approach for interference management is presented. The goal is to improve the systems performance of a macro-cellular network overlaid by femto-cells. In order to improve the time of convergence, a mitigation approach has been introduced, allowing them to have significant gains in terms of throughput for both, macro and femto users. Interesting trade-offs can be studied to compare centralized vs. distributed solutions. In the novel context of small cells distributed solutions to interference management are to be preferred over more complex centralized solutions, but convergence and instability approaches may appear to affect the TD learning schemes, compromising system performances \cite{Galindo}.

\subsubsection{Energy Savings}

Energy savings schemes for wireless cellular systems have been proposed in the past, enabling cells to go into a sleep mode, in which they consume a reduced amount of energy. In order to reduce the energy consumption of the eNBs, we can found several works related to ML techniques. An example of that can be found in \cite{miozzo}, where the authors take advantage of RL to propose a decentralized Q-learning approach to allow energy savings by learning a policy by the iterations with the environment taking into account different aspects over time, such as the daily solar irradiation. Also, in \cite{annaCAMAD}, the authors switch off some underutilized cells during off peak hours. The proposed approach optimizes the number of base stations in dense LTE pico cell deployments in order to maximize the energy saving. For the purpose, they use a combination of Fuzzy Logic, Grey Relational Analysis and Analytic Hierarchy Process tools to trigger the switch off actions, and jointly consider multiple decision inputs for each cell. This last work uses smart decision theory approaches, which though are not able to take advantage of the previous decisions made in the same environment, as in turn does the work proposed in \cite{miozzo}, as a result of the TD learning approach. This allows that the work in \cite{miozzo} offers a more solid solution, considering also past information in the decision. Also for HetNets, we find several works, such as, \cite{es1,es2}, where the authors take advantage of \acp{KPI}s available in the network for the construction of different kind of databases to analyse the potential gains that can be achieved in clustered small cell deployments.

\subsubsection{Cell Outage Compensation}

The literature already offers different works targeting the problem of COC. For this use case RL has been proven as a valid solution since it is a continuous decision making/control problem. In this context a contribution in the area of self-healing has been presented in \cite{jmoysenSH, onireti}, where the authors present a complete solution for the automatic mitigation of the degradation effect of the outage by appropriately adjusting suitable radio parameters of the surrounding cells. The solution consists of optimizing the coverage and capacity of the identified outage zone, by adjusting the gain of the antenna due to the electrical tilt and the downlink transmission power of the surrounding \ac{eNB}s. To implement this approach, the authors propose a \ac{RL} based on actor-critic theory to take advantage of its capability of making online decisions at each \ac{eNB}, and of providing decisions adapting to the evolution of the scenario in terms of mobility of users, shadowing, etc., and of the decisions made by the surrounding nodes to solve the same problem. A COC contribution also based on ML is targeted in \cite{ARSALAN}, where fuzzy logic is proposed as the driving techniques to fill a coverage gap. The authors show performance gains by using different parameters, such as, the power transmission, the antenna tilt, and a combination of the two schemes. These two works are compared in \cite{jmoysenSH} and the work in \cite{jmoysenSH} is proven superior thanks to the ability to learn from the past experience introduced by the RL actor-critic approach.
\subsubsection{Cell Outage Detection}
As we already mentioned, COD aims to autonomously detect cells that are not operating properly due to possible failures. For this kind of problem, anomaly detection algorithms offer an interesting solution that allows to identify outliers measurements, which can be highlighting a hidden problem in the network. Proposals of solutions for this problem can be found in \cite{fedor} and \cite{khabib}. In particular, \cite{fedor} presents a solution based on diffusion maps, by means of clustering schemes, capable of detecting anomalous behaviours generated by a sleeping cell. \cite{khabib} presents a solution based on fuzzy logic for the automatic diagnosis of a troubleshooting system. In order to determine if there is a failure, the authors propose a controller, which receives as an inputs a set of representative KPIs. A similar approach is presented by \cite{RANA}, where the authors present an automated diagnosis model for \ac{UMTS} networks based on Naive Bayesian classifier, and where the model uses both network simulator and real UMTS network measurements. In the context of this king of classifiers , the works in \cite{Barco05,Khanafer08}, also take advantage of NB for automated diagnosis based on different inputs network performances.
The work in \cite{emcoc} addresses both the case of outage and the one where in turn the cell can provide a certain level of service, which though does not allow to fulfil the expected UEs requirements. The approach relies on ensemble methods to train KPIs extracted by human operators to make informed decisions.
In \cite{turkka3}, the authors consider large data sets to identify anomaly behaving base station. They proposed an algorithm consisting of preprocessing, detection and analysis phases. The results show that by using dimensionality reduction and anomaly detection techniques irregularly behaving base stations can be detected in a self-organized manner. In \cite{onireti} data gathered through MDT reports is used for anomaly detection purposes. Furthermore, the works of \cite{Xue14,Zoha14,Chernov14} take advantage of $k$-NN algorithm to propose a self-healing solution, in particular to tackle the fault detection domain.
Finally, in \cite{Alias16}, the authors consider a HetNet and they take advantage of HMM to automatically capture the dynamic's of four different states and probabilistically estimate if there exist a possible failure.

\subsubsection{SON Conflicts Coordination}
As the deployment of stand-alone SON functions is increasing, the number of conflicts and dependencies between them also increases. Hence, an entity has been proposed for the coordination of this kind of conflicts. In this context, current literature includes several works based on ML. In \cite{hafiz} the authors focus on the classification of potential SON conflicts and on discussing the valid tools and procedures to implement a solid self-coordination framework. Q-learning, as a \ac{RL} method, has been proposed in \cite{Berna_coord} to take advantage of experience gained in past decisions, in order to reduce the uncertainty associated with the impact of the SON coordinator decisions when picking an action over another to resolve conflicts. In \cite{lbHo}, the authors use Q-learning to deal with the conflict resolution between two SON instances. Decision trees have been proposed in \cite{policy} to properly adjust \ac{RET} and transmission power. Additionally, in \cite{jmoysenEurasip} the authors provide a functional architecture that can be used to deal with the conflicts generated by the concurrent execution of multiple SON functions. They show that the proposed approach is general enough to model all the SON functions and their derived conflicts. First they introduce these SON functions in the context of the general SON architecture, together with high-level examples of how they may interfere. Second, they define the state and action spaces of the global \ac{MDP} that models the self-optimization procedure of the overall \ac{RAN} segment. Finally, they show that the global self-optimization problem can be decomposed onto as many \acp{subMDP} as \ac{SON} functions.

\subsubsection{Minimization of Drive Tests}

The great majority of literature using the MDT functionality to target MDT use cases, takes advantage of supervised and unsupervised learning techniques to provide different solutions for the different use cases. An example of that can be observed in \cite{chernogorov12,chernogorov13}, where the authors address the QoS estimation by selecting different KPIs and correlating them with common nodes measurements, to establish whether a UE is satisfied with the received QoS. A similar objective is targeted in \cite{jmoysenCAMAD}, however, differently from the previous works, here the authors focus on multi layer heterogeneous networks, so in a more complex and realistic scenario than the traditional macrocell one. In particular, they present an approach, based on regression models, which allows to predict QoS in heterogeneous networks for UEs, independently of the physical location of the UE. This work is extended in \cite{jmoysenISCC} by taking into account the most promising regression models, but also analysing dimensional reduction techniques. By doing \ac{PCA}/\ac{SPCA} on the input features, and promoting solutions in which only a small number of input features capture most of the variance, the number of random variables under consideration is reduced. Based on previous results, in \cite{jmoysenPIMRC, jmoysenHindawi} the same authors define a methodology to build a tool for smart and efficient network planning, based on QoS prediction derived by proper data analysis of UE measurements in the network.

Moreover, the work in \cite{matiasCCO} presents a system based on a fuzzy logic controller to improve network performances by adjusting antenna tilts values in a LTE system. Differently from previous works, the authors consider the use of call traces to identify the level of coverage, overshooting and overlapping problems, which are the inputs to the algorithm. Also, in \cite{matias}, the same authors take advantage of connection traces (signal strength, traffic, and resource utilization measurements) to improve the network infrastructure in terms of spectral efficiency. The proposed method is designed to be integrated in commercial network planning tool. Finally, in \cite{AnaMariaMDT} the authors take advantage of the MDT measurements to build a \ac{REM} by applying spatial interpolation techniques (Bayesian kriging). The REM (Radio Environmental Map) is then used to detect coverage holes and predict the shape of those areas.

\subsubsection{Core Networks}
As we already mentioned in section \ref{sec:son}, the operational aspects of core networks elements can be enhanced through, for example, the automatic configuration of the neighbour cell relations function. In this regard, the idea of applying ML to this function is not new. In \cite{balint} the authors study the benefits of using ML to root-cause analysis of session drops, as well as drop prediction for individual sessions. They present an offline Adaboost and SVM method to create a predictor, which is in charge of eliminating/mitigating the session drops by using real LTE data.

\subsubsection{Virtualized and Software Define Networks}
Also when we go beyond the \ac{RAN} and we focus on the network in general, ML concepts have already been proposed in different works to build cognitive based techniques to operate the network. An example of these proposals is well summarized by~\cite{Clark13}. In this work, a Knowledge Plane is advocated, which would bring many advantages to the networks in terms how the network is operated, automated, optimized and troubleshooted. Conceptually this vision is aligned with different others proposals in other areas, such as the black-box optimization~\cite{Rios2013}, the autonomic self-x architectures~\cite{Derbel2009}, or the work presented in~\cite{Zorzi16}. In this context, the work in~\cite{Mestres16} analyzes the reasons why the vision proposed in~\cite{Clark13} has still not been brought to reality, and the main reason that they find is in the challenges that appear when it comes to autonomously manage a network in a distributed fashion. In particular, the work argues that the emerging trend of centralization in control brought by the novel \ac{SDN} vision, will significantly reduce this complexity and favour the realization of the ML vision in the network. As a result, in~\cite{Mestres16} some initial experimental results based on the vision defined in~\cite{Clark13} are brought into reality in the context of a \ac{SDN} based architecture.
Further work in this area is carried put in the context of different European H2020 projects~\cite{COGNET}. The work in~\cite{Yahia17} presents a novel cognitive management architecture that manages multiple use cases, like the Service Level Agreement (SLA) and the Mobility Quality Predictor. Both use cases are tackled using machine learning approaches, the Long Short Term Memory, and a per user bandwidth predictor. The work in~\cite{Bendriss17} implements SLA through ML approaches. It uses an \ac{ANN} for evaluation of cognitive SLA enforcement of networking services involving Virtualized Network Functions and \ac{SDN} controllers.

\section{Challenges for future works}
\label{sec:challenges}
In this section, we focus on some open challenges that still need to be addressed when it comes to making ML based network management a reality.

\subsection{Real data}
It is possible to find databases related to signals and coverage data \cite{opensignal, D4D}, by using/designing applications that collect information such as \ac{RSRP} and \ac{RSRQ}. However, it is not easy to find contributions analysing real network management data. Some work can be found in the context of 3G networks, but currently, in the context of 4G networks it is very hard to find works considering real data \cite{european,laner}.
These works though do not take into account the data analysing them through ML techniques, to extract experience from them. We consider that it is extremely important for this research line to get to the next level, to get access to operator's network data.
An alternative to real data, could be to sniff data from unencrypted LTE control channels, as we have shown in \cite{Trinh17} or to use a high-fidelity network simulator ns-3 \ac{LTE}/\ac{LENA} module, to generate realistic data \cite{lena}. This simulator has been built around industrial \acf{API} defined by the small cell forum and offers high-fidelity models from \ac{MAC} to application layers. It has also been designed with the requirement to simulate tens of eNBs and hundreds of UEs, and to specifically test \ac{RRM} and SON algorithms. Consequently it could be a very useful tool to build realistic scenarios based on information available in public databases, generate data to analyse, build algorithms based on this analysis and close the loop on the simulator to test the designed algorithms. In this context, it is also hard to find contributions where ML approaches are used not only in network simulators, but in real networking products. In general, it seems vendors are reluctant to test algorithms whose behaviour is not predictable.
An important research line is then how to find or generate meaningful network data, and find patterns in them to understand aspects that should be optimized in the network.

This research line, additionally, faces important privacy and confidentiality issues. It is important to ensure that the data that is used is properly anonymized. As mentioned in section \ref{sec:MLinSON}, data come from different sources of the network, but can also be offered by third parties, e.g., data
generated by the user, open data, sensor data, among others. Therefore, to come up with a unified
privacy policy is extremely challenging at security and privacy levels, due to the variety and the granularity of the
data. If we add to this the speed at which data are created and need to be analysed, the security challenges are huge. In this context, big data is changing the security analytics, where robust and scalable privacy
preserving mining algorithms are critical to ensure that the most sensitive private data is secure. As a result, privacy-preserving data mining is a challenging research line that has
to be investigated. In particular, in order to guarantee privacy protection, it is important to define the privacy requirements taking into account the lifecycle of the analytics. For example, in the data collection phase, it is important to identify the personal data  needed for processing. The idea is to extract only the needed data for the specific purpose. Aggregated information can also be used instead of personal data. In this context, one of the most relevant techniques is anonymization, which is the process of modifying personal data in such a way that no identification is possible. Regarding the data analysis phase, different privacy models are available in the context of big data analytics, where two of the most important families are: K-anonymity and differential privacy. A review in more detail of the aforementioned methods can be found in \cite{enisa}. Moreover, in order to protect personal data in databases (data sotrage phase), \emph{encryption} is a fundamental security technique, which transforms data in a way that only authorized parties can read it. For more information, the reader is referred to \cite{enisa,Khairulliza16}.

\subsection{Big Data and Deep Learning}
Deep learning is a new trend in \ac{ML} that allows computer systems to improve with experience and data. It achieves great power and flexibility to operate in complicated real-world environments, by learning to represent the world through a nested hierarchy of concepts. The ML algorithms that we have reviewed in this paper have a strong dependency on the features on which the algorithms are applied. Based on that, much effort has been devoted to design ML algorithms that yield to useful representations. This is known as \emph{representation learning}, and deep learning is one way of learning representations \cite{Bengio13,DL2016}. The main representation of deep learning is through a Multilayer Perceptron, which is a multilayer neural network function mapping some sets of input values to output values. Each layer of this representation learns a hierarchy of the output. Deep Learning has the ability to do successful training from the bottom layers to the higher ones. This is done by applying computational models that are composed of multiple levels of representation and abstraction that help make sense of data.

Historically, DL has become more useful as the amount of training data has started increasing (big data). Also, the research on deep learning has benefited from the increase of computer infrastructure at both hardware and software levels. All this has made that deep learning has solved increasingly complicated applications with increasing time and accuracy. The potential of these improved techniques in the area of NM, in case the big data associated to the management of the complex 4G, 5G network ecosystem is available, is still to be evaluated and open for research.

\subsection{Theoretical research}
With respect to online control decision problems that allow to continuously take RRM/SON decisions, we are aware of some approaches, which take advantage of reinforcement learning to solve this problem. The current approach is to use single agent algorithms and extend them to multi-agent settings. However, this kind of algorithms require a considerable amount of time before finding a solution, and it increases with the state and action spaces. So, reinforcement learning approaches dealing with this issues have to be investigated. Moreover, no proof of convergence is available demonstrating that this approach actually reaches meaningful conclusions.
Even though ML literature offers different algorithms that can find interesting solutions (e.g. NashQ \cite{nashQ}), the space of possible solutions is so big that this kind of approaches is not feasible to be used in a realistic network where the time constraints of RRM/SON problem have to be met. So, more research in the area of multi-agent systems, which are also compatible with real network requirements need to be investigated.

In the context of data analytics, it is well known that the analysis of the data requires a substantial amount of "\textit{black art}", and consequently it requires the availability in the research groups of multi-disciplinary researcher profiles knowledgeable of information technology, computer science, and telecom engineering, to  properly optimize the network accordingly. In this context, ML trends, like deep learning can be very useful, however little work can be found applying these new promising techniques to network management \cite{zorzi}.

\subsection{Network management of multi-technologies networks and of future New Radio}
Autonomous network management of multi-technology networks, where heterogeneous networks including different \ac{RAT}, or different layers of the network are coexisting, e.g., Wi-Fi, mmWave, mobile network layer, transport layer, among others, is still immature. However, these scenarios will tend to emerge with the advent of the unlicensed spectrum paradigm and with technologies such as LAA or New Radio (NR), the new radio access technology for 5G, which is currently under definition in \ac{3GPP}. NR, in particular, will be defined to work over a wide range of spectrum opportunities, ranging from sub 6 GHz and up to mmWave spectrum, and under multiple spectrum paradigms, such as licensed, unlicensed and shared. The opportunities of autonomous network management in this area are huge. ML has still not been exploited to handle these networks with intelligence and self-awareness. In particular, the management of densified and heterogeneous, in both technologies and layers, architectures, requires the evolution of complex SON concepts, which have traditionally been designed and standardized for LTE based networks. Also, self-organization in the context of NR technology is still to be completely defined. Before reaching this vision, multiple challenges need to be addressed, e.g. the self-coordination problem and the solution of conflicts among SON functions executed in different nodes, or networks, which put the network at risk of instability, or the most appropriate location of SON functions and algorithms, to solve properly the distributed vs. centralized SON implementation issue. Many aspects have to be considered when locating and designing a SON function, e.g. response time, complexity, size of databases, computational capability of nodes, etc. Centralized (i.e. a large number of cells is involved), distributed (approx. 2 cells are involved, coordinating through X2) and local (only one cell is involved) implementations of SON functions have been proposed. No architecture can be claimed superior to the other. The growing complexity, dynamicity, and heterogeneity of 5G networks will substantially increase the number of scenarios to solve. So, there is the need for exploiting their complementarity by virtualizing and dynamically deploying them.

\subsection{Network management of novel softwarized and virtualized architectures}
To benefit of all the opportunities offered by centralized, distributed and local implementations, and towards the need of virtualizing resources in order to reduce network costs, while meeting the stringent new service verticals' requirements, there is the need to further study autonomous NFV and SDN architecture, where end-to-end SON functions, aimed at tackling the main radio access and backhauling challenges of extremely dense deployments, are virtualized and run over generic purpose hardware. This infrastructure is to be managed by an orchestrator entity (in coordination with the corresponding virtual network function and virtual infrastructure managers), as proposed in \ac{ETSI} architecture. This orchestrator or SDN controller is the brain of the network and needs the ability to adapt to ever-changing conditions. The network should not only react to failures, but adapt to the demand, predict it, based on data analytics, and facilitate in this way the task of the network management. Research on deep reinforcement learning implementations of the orchestrator will allow the controller to self-learn after every decision. Automation will require also all the advancements of the Information Technology sector, with increased computational capacity, more CPUs and memory space. However, future orchestrators will need to handle a huge amount of data and learn from them through novel deep learning approaches, the smart network management decisions. This research line is still highly immature and requires a lot of efforts.

\section{Conclusions}
\label{sec:conclusions}
In this work we have motivated the need for ML to be considered as a crucial and inevitable tool in order to address automation, self-awareness and self-organization in current and future mobile networks. The SON features have been considered fundamental in LTE definition and have been introduced in this technology since its very beginning in Release 8. We believe that this need of automation will be further enhanced with the expected complexity that future 5G network management will have to handle. On top of that, we have shown that current cellular networks, already generate a huge amount of data that if properly stored and managed could bring new insights in how the networks work and offer new challenges for improving network management taking into account the experience that can be gained from these data. We have reviewed the main taxonomy of machine learning and the novel trends that could make this exploitation of data to gain insight of the network a reality. Also, we have discussed open data options, as much as alternatives to get data from the networks, which otherwise are not made available to the academic community. With this motivations in mind, we have started by reviewing the main concepts and taxonomy of SON, network management and ML, and we have reviewed significant academic literature in the area of network management, focusing only on solutions based on ML.
The work has reviewed the status of this exciting research line, while at the same time highlighting open challenges that we need to deal with in order to make future autonomous network management a reality.

\section{COMPETING INTERESTS}
The authors declare that there is no conflict of interest regarding the publication of this paper.

\section{ACKNOWLEDGMENT}
The research leading to these results has received funding from the Spanish Ministry of Economy and Competitiveness under grant TEC2014-60491-R (Project 5GNORM). This work also was supported by the Spanish National Science Council and ERFD funds under Project TEC2014-60258-C2-2-R.

\bibliographystyle{IEEEtran}
\bibliography{mybibtex}

\begin{thebibliography}{100}
\providecommand{\url}[1]{#1}
\csname url@samestyle\endcsname
\providecommand{\newblock}{\relax}
\providecommand{\bibinfo}[2]{#2}
\providecommand{\BIBentrySTDinterwordspacing}{\spaceskip=0pt\relax}
\providecommand{\BIBentryALTinterwordstretchfactor}{4}
\providecommand{\BIBentryALTinterwordspacing}{\spaceskip=\fontdimen2\font plus
\BIBentryALTinterwordstretchfactor\fontdimen3\font minus
  \fontdimen4\font\relax}
\providecommand{\BIBforeignlanguage}[2]{{%
\expandafter\ifx\csname l@#1\endcsname\relax
\typeout{** WARNING: IEEEtran.bst: No hyphenation pattern has been}%
\typeout{** loaded for the language `#1'. Using the pattern for}%
\typeout{** the default language instead.}%
\else
\language=\csname l@#1\endcsname
\fi
#2}}
\providecommand{\BIBdecl}{\relax}
\BIBdecl

\bibitem{Gandalf}
\BIBentryALTinterwordspacing
{Gandalf}. {Monitoring and self-tuning of RRM parameters in a multi-system
  network}. [Online]. Available: \url{{http://www.celtic-initiative.org/}}
\BIBentrySTDinterwordspacing

\bibitem{SOCRATES}
\BIBentryALTinterwordspacing
{SOCRATES}. {Self-Optimisation and self-ConfiguRATion in wirelEss networkS}.
  [Online]. Available: \url{{http://www.fp7-socrates.org/}}
\BIBentrySTDinterwordspacing

\bibitem{SEMAFOUR}
\BIBentryALTinterwordspacing
{SEMAFOUR}. {Self-Management for Unified Heterogeneous Radio Access Networks}.
  [Online]. Available: \url{{http://fp7-semafour.eu/}}
\BIBentrySTDinterwordspacing

\bibitem{SESAME}
\BIBentryALTinterwordspacing
{SESAME}. {Small Cells Coordination for Multi-tenancy and Edge services}.
  [Online]. Available: \url{https://5g-ppp.eu/sesame/}
\BIBentrySTDinterwordspacing

\bibitem{SELFNET}
\BIBentryALTinterwordspacing
{SELFNET}. {Framework for Self-Organised Network Management in Virtualised and
  Software Defined Networks}. [Online]. Available:
  \url{{http://www.cognet.5g-ppp.eu/cognet-in-5gpp/}}
\BIBentrySTDinterwordspacing

\bibitem{COGNET}
\BIBentryALTinterwordspacing
{COGNET}. {Cognitive Networks}. [Online]. Available: \url{{https://5g-ppp.eu/}}
\BIBentrySTDinterwordspacing

\bibitem{survey}
{O. Aliu, A. Imran, M. Imran, B. Evans}, ``{A Survey of Self Organization in
  Future Cellular Networks},'' \emph{IEEE Communications Surveys $\&$
  Tutorials}, no.~99, pp. 1--26, 2012.

\bibitem{sistelbanda}
\BIBentryALTinterwordspacing
{Sistelbanda}. {SN4G {SON}}. [Online]. Available:
  \url{{http://sistelbanda.es/}}
\BIBentrySTDinterwordspacing

\bibitem{qualcom}
\BIBentryALTinterwordspacing
Qualcomm. {Self managing and enabling seamless roaming}. [Online]. Available:
  \url{{https://www.qualcomm.com/videos/qualcomm-wi-fi-son}}
\BIBentrySTDinterwordspacing

\bibitem{huawei}
\BIBentryALTinterwordspacing
Huawei. {Huawei's Innovative Single {SON} Solution}. [Online]. Available:
  \url{{http://www1.huawei.com/en/solutions/broader-smarter/}}
\BIBentrySTDinterwordspacing

\bibitem{airhop}
\BIBentryALTinterwordspacing
A.~communications. {Powering 4G networks}. [Online]. Available:
  \url{{http://www.airhopcomm.com}}
\BIBentrySTDinterwordspacing

\bibitem{cellwize}
\BIBentryALTinterwordspacing
Cellwize. Driving value through {SON}. [Online]. Available:
  \url{{http://www.cellwize.com}}
\BIBentrySTDinterwordspacing

\bibitem{aviat}
\BIBentryALTinterwordspacing
{AVIAT}. {Wireless products for small cell applications}. [Online]. Available:
  \url{{https://startupgenome.co/aviat-networks}}
\BIBentrySTDinterwordspacing

\bibitem{smallcellForum}
\BIBentryALTinterwordspacing
S.~C. Forum. Small cell industry award. [Online]. Available:
  \url{{http://www.smallcellforum.org/events/awards}}
\BIBentrySTDinterwordspacing

\bibitem{3GPPwork}
3GPP, ``{Technical Specification Group Services and System Aspects;
  Telecommunication Management; Self-Organizing Networks ({SON}); Concepts and
  requirements (Release 13)},'' 3GPP, TS 32.500, v13.0.0, 2016.

\bibitem{challengesSON}
{A. Imran, A. Zoha, A. Abu-Dayya}, ``{Challenges in 5G: How to Empower {SON}
  with Big Data for Enabling 5G},'' \emph{IEEE Network}, vol.~28, no.~6, pp.
  27--33, 2014.

\bibitem{nsn}
NOKIA, ``{Intelligent Self-Organizing Networks: Increased automation for higher
  network performance with lower costs},'' \emph{White paper}, 2013.

\bibitem{bDemp}
{N. Baldo, L. Giupponi, J. Mangues-Bafalluy}, ``{Big Data Empowered Self
  Organized Networks},'' \emph{20th IEEE European Wireless}, 2014.

\bibitem{densification}
{B. Romanous, N. Bitar, A. Imran, H. Refai}, ``{Network Densification:
  Challenges and Opportunities in enabling 5G},'' \emph{20th {IEEE}
  International Workshop on Computer Aided Modelling and Design of
  Communication Links and Networks (CAMAD, Special Session: Dense and Elastic
  Big Data Empowered {SON}}, pp. 129--134, 2015.

\bibitem{dressler}
{D. Falko}, ``{A Study of Self-Organization Mechanisms in Ad Hoc and Sensor
  Networks},'' \emph{Elseiver Computer Communications}, pp. 3018--3019, 2008.

\bibitem{SDN-survey}
{D. Kreutz, F. M. V. Ramos, P. E. Verissimo, C. E. Rothenberg, S. Azodolmolky,
  S. Uhlig}, ``{Software-Defined Networking: A Comprehensive Survey},''
  \emph{Proceedings of the IEEE}, vol. 103, no.~1, pp. 14--76, Jan 2015.

\bibitem{3GPPUSES}
3GPP, ``{Evolved Universal Terrestrial Radio Access Network (E-UTRAN); Self
  configuring and self optimizing network uses case and solutions (Release
  9)},'' 3GPP, TR T36.902, v9.2.0, 2010.

\bibitem{TS32.500}
------, ``{Telecommunication management; Self-Organizing Networks ({SON});
  Concepts and requirements (Release 11)},'' 3GPP, TS 32.500, v11.1.0, 2011.

\bibitem{TS32.501}
{3GPP}, ``{Self Configuration of Network Elements; Concepts and Requirements,
  Release 9},'' 3GPP, TS 32.501, 2010.

\bibitem{TS32.502}
3GPP, ``{Self-configuration of network elements Integration Reference Point
  (IRP); Information Service (IS) (Release 10)},'' 3GPP, TS 32.502, v10.1.0,
  2010.

\bibitem{TS32.503}
------, ``{Self-Configuration of Network Elements Integration Reference Point
  (IRP); Common Object Request Broker Architecture (CORBA) Solution Set
  (SS)(Release 8)},'' 3GPP, TR 32.503, 2008.

\bibitem{TS32.531}
------, ``{Software Management (SWM); Software and Integration Reference Point
  (IRP) Requirements (Release 10)},'' 3GPP, TR 32.531, 2010.

\bibitem{TS32.532}
------, ``{Software management Integration Reference Point (IRP); Information
  Service (IS) (Release 8)},'' 3GPP, TR 32.532, 2009.

\bibitem{TS32.533}
------, ``{Software Management (SWM) Integration Reference Point (IRP): Common
  Object Request Broker Architecture (CORBA) Solution Set (SS) (Release 9)},''
  3GPP, Tech. Rep. 32.533, 2009.

\bibitem{TS32.761}
------, Tech. Rep.

\bibitem{TS32.762}
------, ``{Evolved Universal Terrestrial Radio Access Network (E-UTRAN) Network
  Resource Model (NRM) Integration Reference Point (IRP); Information Service
  (IS) (Release 11)},'' 3GPP, TS 32.762, v11.7.0, 2014.

\bibitem{TS32.763}
------, ``{Evolved Universal Terrestrial Radio Access Network (E-UTRAN) Network
  Resource Model (NRM) Integration Reference Point (IRP): CORBA solution set
  (Release 9)},'' 3GPP, TR 32.763, 2009.

\bibitem{TS32.765}
------, ``{Evolved Universal Terrestrial Radio Access Netwrok (E-UTRAN) Network
  Resource Model (NRM) Integration Reference Point (IRP): eXtensible Markup
  Language (XML) definitions (Release 9)},'' 3GPP, TR 32.765, v9.6.0, 2004.

\bibitem{TS32.821}
------, ``{Study of Self-Organizing Networks ({SON}) related Operations,
  Administration and Maintenance (OAM) for Home Node B (HNB) (Release 9)},''
  3GPP, TR 32.821, v9.0.0, 2004.

\bibitem{TR32.823}
------, ``{Telecommunication management; Self-Organizing Networks ({SON});
  Study on self-healing (Release 9)},'' 3GPP, TR 32.823, 2010.

\bibitem{TS32.425}
{3GPP}, ``{Performance measurements Evolved Universal Terrestrial Radio Access
  Network (E-UTRAN)(Release 10)},'' {3GPP}, TS 32.425, v10.6.0, 2012.

\bibitem{TS25.413}
3GPP, ``{Universal Mobile Telecommunications System (UMTS); UTRAN Iu Interface
  RANAP Signalling},'' 3GPP, TS 25.413, v3.4.0, 1999.

\bibitem{TS36.300}
------, ``{Radio Access (E-UTRA) and Evolved Universal Terrestrial Radio Access
  Network (E-UTRAN); Overall description; Stage 2 (Release 12)},'' 3GPP, TS
  36.300, 2015.

\bibitem{TS36.413}
------, ``{{LTE}; Evolved Universal Terrestrial Radio Access Network (E-UTRAN);
  S1 Application Protocol (S1AP) (Release 9)},'' 3GPP, TS 36.413, v9.4.0, 2010.

\bibitem{TS36.423}
{3GPP}, ``{Technical Specification Group Radio Access Network; Evolved
  Universal Terrestrial Radio Access Network (EUTRAN); X2 Application Protocol
  (X2AP) (Release 11)},'' {3GPP}, TS 36.423, v10.7.0, 2013.

\bibitem{TS32.522}
3GPP, ``{Self-Organizing Networks ({SON}) Policy Network Resource Model (NRM)
  Integration Reference Point (IRP); Information Service(IS), 3rd Generation
  Partnership Project (3GPP), (Release 10)},'' 3GPP, TS 32.522, 2011.

\bibitem{TS32.526}
------, ``{Self-Organizing Networks ({SON}): Policy Network Resource Model
  (NRM) Integration Reference Point (IRP); Information Service(IS); Solution
  Set (SS) definitions, (Release 10)},'' 3GPP, TS 32.526, 2011.

\bibitem{TS32.766}
------, ``{Evolved Universal Terrestrial Radio Access Network (E-UTRAN);
  Network Resource Model (NRM) Integration Reference Point (IRP); Solution
  Set(SS) definitions, (Release 10)},'' 3GPP, TS 32.766, 2011.

\bibitem{TS32.541}
------, ``{Telecommunication management; Self-Organizing Networks ({SON});
  Self-healing concepts and requirements (Release 12)},'' 3GPP, TS 32.541,
  v12.0.0, 2014.

\bibitem{TR32.826}
------, ``{Technical Specification Group Services and System Aspects;
  Telecommunication Management; Study on Energy Savings Management (ESM)
  (Release 10)},'' 3GPP, TR 32.826, 2010.

\bibitem{TR32.834}
------, ``{Technical Specification Group Services and System Aspects; Study on
  Operations, Administration and Maintenance (OAM) aspects of
  inter-Radio-Access-Technology (RAT) energy saving (Release 11)},'' 3GPP, TR
  32.834, 2012.

\bibitem{TS32.551}
------, ``{Universal Mobile Telecommunications System (UMTS); {LTE};
  Telecommunication management; Energy Saving Management (ESM); Concepts and
  requirements (Release 10)},'' 3GPP, TS 32.551, v10.1.0, 2011.

\bibitem{TS36.331}
------, ``{Technical Specification Group Radio Access Network; Evolved
  Universal Terrestrial Radio Access (E-UTRA); Radio Resource Control (RRC);
  Protocol Specification (Release 10)},'' 3GPP, TS 36.331, v10.7.0, 2010.

\bibitem{TS32.405}
------, ``{Performance Management (PM); Performance Measurements UTRAN (Release
  9)},'' 3GPP, TS 32.405, 2009.

\bibitem{TR32.511}
------, ``{Automatic Neighbor Relation (ANR) management, Concepts and
  Requirements, (Release 11)},'' 3GPP, TR 32.511, 2011.

\bibitem{TS32.521}
{3GPP}, ``{Technical Specification Group Services and System Aspects;
  Telecomunications Management; Self-Organizing Networks ({SON}) Policy Network
  Resource Model (NRM) Integration Reference Point (IRP)(Release 10)},''
  {3GPP}, TR 32.521, v10.2.0, 2009.

\bibitem{TS32.642}
3GPP, ``{Configuration Management (CM); UTRAN network resources Integration
  Reference Point (IRP); Network Resource Model (NRM)(Release 9)},'' 3GPP, TS
  322.642, 2009.

\bibitem{S5-122330}
------, ``{TSG SA WG5 (Telecom Management) Meeting 85; Study of implementation
  alternative for {SON} coordination},'' 3GPP, Tech. Rep. S5-122330, 2012.

\bibitem{TS32.646}
------, ``{Configuration Management (CM); UTRAN network resources Integration
  Reference Point (IRP); Solution Set (Release 10)},'' 3GPP, Tech. Rep. 32.646,
  v10.3.0, 2011.

\bibitem{TS25.331}
------, ``{Radio Resource Control (RRC); Protocol specification (Release
  10)},'' 3GPP, TS 25.331, v10.6.0, 2011.

\bibitem{TS25.401}
------, ``{UTRAN overall description (Release 10)},'' 3GPP, TS 25.401, 2011.

\bibitem{TS25.410}
------, ``{UTRAN Iu Interface: General Aspects and Principles (Release 10)},''
  3GPP, TS 25.410, 2011.

\bibitem{TS25.423}
------, ``{UTRAN Iur interface Radio Network Subsystem Application Part (RNSAP)
  signalling (Release 9)}, institution = {3GPP},'' TS 25.423, 2010.

\bibitem{TS32.836}
------, ``{Study on Network Management (NM) centralized Coverage and Capacity
  Optimization (CCO) Self-Organizing Networks ({SON}) function},'' 3GPP, TS
  32.836, 2013.

\bibitem{TS32.103}
------, ``{Integration Reference Point (IRP) overview and usage guide (Release
  10)},'' 3GPP, TS 32.103, v10.0.0, 2011.

\bibitem{TS28.627}
------, ``{Self-Organizing Networks ({SON}) Policy Network Resource Model (NRM)
  Integration Reference Point (IRP); Requirements (Release 11)},'' 3GPP, TS
  28.627, v11.1.0, 2012.

\bibitem{TS28.628}
------, ``{Self-Organizing Networks ({SON}) Policy Network Resource Model (NRM)
  Integration Reference Point (IRP); Information Service (IS) (Release 11)},''
  3GPP, TS 28.628, v11.4.0, 2014.

\bibitem{TS28.658}
------, ``{Evolved Universal Terrestrial Radio Access Network (E-UTRAN) Network
  Resource Model (NRM) Integration Reference Point (IRP); Information Service
  (IS) (Release 11)},'' 3GPP, TS 28.658, v11.0.0, 2012.

\bibitem{TS28.659}
------, ``{Evolved Universal Terrestrial Radio Access Network (E-UTRAN) Network
  Resource Model (NRM) Integration Reference Point (IRP); Solution Set (SS)
  definitions (Release 11)},'' 3GPP, TS 28.659, 2012.

\bibitem{TS32.508}
------, ``{Gap analysis between 3GPP SA5 specifications and NGMN Top
  Operational Efficiency (OPE) Recommendations (Release 12)},'' 3GPP, TS
  32.508, 2013.

\bibitem{TS32.509}
------, ``{Data formats for multi-vendor plug and play eNodeB connection to the
  network (Release 12)},'' 3GPP, TS 32.509, 2013.

\bibitem{TR32.838}
------, ``{Gap analysis between 3GPP SA5 specifications and NGMN Top
  Operational Efficiency (OPE) Recommendations (Release 12)},'' 3GPP, TR
  32.838, 2013.

\bibitem{TR32.835}
------, ``{Study of heterogeneous networks management (Release 12)},'' 3GPP, TR
  32.835, v12.0.0, 2014.

\bibitem{TR32.851}
------, ``{Study on Operations, Administration and Maintenance (OAM) aspects of
  Network Sharing (Release 12)},'' 3GPP, TR 32.851, v12.1.0, 2013.

\bibitem{TR37.822}
------, ``{Study on Next Generation Self-Optimizing Network ({SON}) for UTRAN
  and E-UTRAN (Release 12) },'' 3GPP, TR 37.822, 2013.

\bibitem{TR36.887}
------, ``{Study on Energy Saving Enhancement for E-UTRAN (Release 12)},''
  3GPP, TR 36.887, 2013.

\bibitem{TR32.860}
------, ``{Study on Enhancements of OAM aspects of distributed Self-Organizing
  Networks ({SON}) functions, Release 12},'' 3GPP, TR 32.860, 2014.

\bibitem{rel14}
------. Overview of 3gpp release 14.

\bibitem{ran14}
\BIBentryALTinterwordspacing
------. (2016) {RAN Evolution of {LTE} in Release 14}. [Online]. Available:
  \url{{http://www.3gpp.org/news-events/3gpp-news/1768-ran_rel14}}
\BIBentrySTDinterwordspacing

\bibitem{TR30.818}
------, ``{Technical Specification Group Services and System Aspects;
  Telecommunication Management; Project scheduling and open issues for SA5,
  (Release 8)},'' 3GPP, Tech. Rep. 30.818, v8.0.8, 2010.

\bibitem{anrpci}
{M. Amirijoo, P. Frenger, F. Gunnarsson}, ``{Neighbor cell relation list and
  physical cell identity self-organization in {LTE}},'' \emph{ICC
  Workshops-IEEE International Conference on Communications Workshops}, pp.
  31--41, March 2008.

\bibitem{parodi}
{F, Parodi, M, Kylvaja, G, Alford, J, Li and J, Pradas}, ``{An Automatic
  Procedure for Neighbor Cell List Definition in Cellular Networks},''
  \emph{IEEE International Symposium on a World of Wireless, Mobile and
  Multimedia Networks (WoWMoM)}, pp. 1--6, June 2007.

\bibitem{TR36.902}
3GPP, ``{Technical Specification Group Radio Access Network, Evolved Universal
  Terrestrial Radio Access Network (E-UTRAN), Self configuring and Self
  optimizing Network ({SON}) Uses Case and Solutions (Release 9)},'' 3GPP, TR
  36.902, v9.3.1, 2011.

\bibitem{ngmn14}
{F. Parodi, M. Kylvaja, G. Alford, J. Li, J. Pradas}, ``{Recommended Practices
  for Multi-vendor {SON} Deployment},'' \emph{D2- NGMN Aliance V1.0}, January
  2014.

\bibitem{TS36.213}
3GPP, ``{Evolved Universal Terrestrial Radio Access (E-UTRA); Physical layer
  procedures},'' 3GPP, TS 36.213, v10.2.0, 2011.

\bibitem{4gngmn}
NGMN, ``{Recommended Practices for Multi-vendor {SON} Deployment},'' \emph{NGMN
  P-SmallCell Work Stream 2 (WS2) first deliverable}, 2014.

\bibitem{36.300}
3GPP, ``{Evolved Universal Terrestrial Radio Access (E-UTRA) and Evolved
  Universal Terrestrial Radio Access{E-UTRAN}; Overall description; Stage 2},''
  3GPP, TS 36.300 v12.6.0, 2015.

\bibitem{S5-460036}
------, ``{SON} self-healing management,'' 3GPP, Tech. Rep. S5-460036, 2011.

\bibitem{son}
{Seppo H\"{a}m\"{a}l\"{a}inen, Henning Sanneck, Cinzia Sartori}, \emph{{{LTE}
  Self-Organising Networks ({SON}): Network Management Automation for
  Operational Efficiency}}.\hskip 1em plus 0.5em minus 0.4em\relax John Wiley
  and Sons, 2012.

\bibitem{schmelz}
{L. C. Schmelz, M. Amirijoo, A. Eisenblaetter, R. Litjens, M. Neuland, J.
  Turk}, ``{A Coordination Framework for Self-Organisation in {LTE}
  Networks},'' \emph{IEEE International Symposium on Integrated Network
  Management}, 2011.

\bibitem{altmanCoord}
{Z. Altman, M.Amirijoo, F.Gunnarson, H.Hoffmann, I.Z. Kov\'{a}cs, D. Laselva,
  B. Sas, K. Spaey, A. Tall, H. Van Den Berg, K. Zetterberg}, ``{On design
  principles for self-organizing network functions},'' \emph{11th International
  Symposium on Wireless Communications Systems (ISWCS)}, pp. 454--459, 2014.

\bibitem{TR36.805}
3GPP, ``{Study on Minimization of drive-tests in Next Generation Networks},''
  3GPP, TR 36.805, 2009.

\bibitem{MDT3GPP}
{J. Johansson, W. A. Hapsari, S. Kelley, G. Bodog}, ``{Minimization of drive
  tests in {3GPP} (Release 11)},'' \emph{IEEE Communications Magazine},
  November 2012.

\bibitem{nokiacore}
\BIBentryALTinterwordspacing
{Nokia Siemens Networks}. (2012) {Automating Core Network Management, Mobile
  World Congress}. [Online]. Available:
  \url{{https://networks.nokia.com/solutions/cloud-native-core-network}}
\BIBentrySTDinterwordspacing

\bibitem{survey-NFV}
{A. Blenk, A. Basta, M. Reisslein, W. Kellerer}, ``{Survey on Network
  Virtualization Hypervisors for Software Defined Networking},'' \emph{IEEE
  Communications Surveys Tutorials}, vol.~18, no.~1, pp. 655--685, 2016.

\bibitem{caruana}
{R. Caruana, A. Niculescu-Mizil}, ``{An empirical comparison of supervised
  learning algorithms},'' \emph{23rd International Conference on Machine
  Learning}, 2006.

\bibitem{glm}
{P. McCullagh, J. Nelder}, ``{Generalized Linear Models, Second Edition},''
  \emph{Boca Raton: Chapman and Hall}, 1989.

\bibitem{nb}
{H. Zhang}, ``{The Optimality of Naive Bayes},'' \emph{17th International
  Florida Artificial Intelligence Research Society Conference}, 2004.

\bibitem{statistical}
{V. Vapnik}, ``{An Overview of Statistical Learning Theory},'' \emph{IEEE
  Transactions on Neural Networks}, vol.~10, no.~5, 1999.

\bibitem{smola}
{A. J. Smola, B. Scholkopf}, ``{A tutorial on support vector regression},''
  \emph{Statistics and Computing}, pp. 199--222, 2004.

\bibitem{andreas}
{A. M\"{u}ller}, ``{Kernel Approximations for Efficient SVMs (and other feature
  extraction methods)},'' \emph{Data Science \& Machine Learning Resources},
  2012.

\bibitem{bishop}
{Bishop M.C.}, ``{Pattern recognition and machine learning},'' \emph{Busines
  Dia, Llc. Springer Science}, 2006.

\bibitem{idt}
{J. R. Quinlan}, ``{Induction of Decision Trees. Machine Learning},''
  \emph{Kluwer Academic Publishers}, pp. 81--106, 1986.

\bibitem{dt}
{L. Rokach, O. Maimon}, ``{Data Mining with Decision Trees: Theory and
  Applications},'' \emph{Series in Machine Perception and Artificial
  Intelligence}, 2008.

\bibitem{em}
{D. Opitz, R. Maclin}, ``{Popular Ensemble Methods: An Empirical Study},''
  \emph{Journal of Artificial Intelligence Research 11}, pp. 169--198, 1999.

\bibitem{em2}
{L. Rokach}, ``Ensemble-based classifiers,'' \emph{{Artificial Intelligence}},
  pp. 1--39, 2010.

\bibitem{em3}
{T. G. Dietterich}, ``{An Experimental Comparison of Three Methods for
  Constructing Ensembles of Decision Trees: Bagging, Boosting and
  Randomization},'' \emph{Machine Learning}, vol.~40, pp. 139--157, 2000.

\bibitem{mlresearch}
------, ``{Machine-Learning Research: Four Current Directions},''
  \emph{American Association for Artificial Intelligence}, 1997.

\bibitem{freundSchapire}
{Y. Freund, R. E. Schapire}, ``{A Decision-Theoretic Generalization of On-Line
  Learning and Application to Boosting},'' \emph{2nd European Conference on
  Machine Learning}, pp. 148--156, 1995.

\bibitem{ul}
{G. Zoubin}, ``{Unsupervised Learning},'' \emph{{Gatsby Computational
  Neuroscience Unit University College London, UK}}, 2004.

\bibitem{kmeans}
{W. H. Press, S. A. Teukolsky, W. T. Vetterling, B. P. Flannery}, ``{Section
  16.1. Gaussian Mixture Models and k-Means Clustering},'' \emph{The Art of
  Scientific Computing (3rd ed.); York: Cambridge University}, 2007.

\bibitem{som}
{A. Ultsch, H. P. Siemon}, ``{Kohonen's Self Organizing Feature Maps for
  Exploratory Data Analysis},'' \emph{International Neural Network Conference},
  1990.

\bibitem{hc}
{T. Hastie, R. Tibshirani, J. Friedman}, \emph{{The Elements of Statistical
  Learning (2nd ed.)}}.\hskip 1em plus 0.5em minus 0.4em\relax New York:
  Springer, 2009.

\bibitem{fuzzyCmeans}
{R. Nock, F. Nielsen}, ``{On Weighting Clustering},'' \emph{IEEE Trans on
  Pattern Analysis and Machine Intelligence}, vol.~28, pp. 1--13, 2006.

\bibitem{jmoysenCAMAD}
{J. Moysen, N. Baldo, L. Giupponi, J. Mangues-Bafalluy}, ``{Predicting QoS in
  {LTE} HetNets based on location-independent UE measurement},'' \emph{20th
  IEEE International Workshop on Computer Aided Modelling and Design of
  Communication Links and Networks, Guildford, UK}, 2015.

\bibitem{jmoysenHindawi}
{J. Moysen, L. Giupponi, J. Mangues-Bafalluy}, ``{A Mobile Network Planning
  tool based on Data Analytics},'' \emph{Mobile Information Systems, Hindawi},
  2017.

\bibitem{dr}
{S. T. Roweis, K. S. Lawrence}, ``{Nonlinear Dimensionality Reduction by
  Locally Linear Embedding},'' \emph{American Association for the Advancement
  of Science}, pp. 2323--2326, 2000.

\bibitem{pca}
{S. T. Roweis}, ``{EM algorithms for {PCA} and {SPCA}},'' \emph{Advances in
  Neural Information Processing Systems, The MIT Press}, 1998.

\bibitem{anomalyDet}
{V. J. Hodge, J. Austin}, ``{A Survey of Outlier Detection Methodologies},''
  \emph{Artificial Intelligence Review}, vol.~22, 2004.

\bibitem{banderaLG}
{I. de la Bandera, R. Barco, P. Mu\~{n}oz, I. Serrano}, ``{Cell Outage
  Detection Based on Handover Statistics},'' \emph{IEEE Communications
  Letters}, vol. 19 (7), 2015.

\bibitem{munozLG}
{P. Mu\~{n}oz, R. Barco, I. Serrano, A. G\'{o}mez Andrades}, ``{Correlation
  Based Time-Series Analysis for Cell Degradation},'' \emph{IEEE Communications
  Letters}, 2016.

\bibitem{schaal}
{S. Schaal and C. Atkeson}, ``{Robot juggling: An implementation of
  memory-based learning},'' \emph{Control Systems Magazine}, 1994.

\bibitem{thrun}
{S. Thrun}, ``{Learning to play the game of chess},'' \emph{Advances in Neural
  Information Processing Systems 7, Cambridge, MA, 1995. The MIT Press.}, 1995.

\bibitem{littman}
{M. L. Littman}, ``{Markov games as a framework for multi-agent reinforcement
  learning},'' \emph{11th International Conference on Machine Learning}, pp.
  157--163, 1994.

\bibitem{panait}
{P. Liviu and L. Sean}, ``{Cooperative Multi-Agent Learning: The State of the
  Art},'' \emph{Autonomous Agents and Multi-Agent Systems}, vol.~11, no.~3, pp.
  387--434, 2005.

\bibitem{suton}
{R. S. Sutton and A. G. Barto}, \emph{{Reinforcement Learning: An
  Introduction}}.\hskip 1em plus 0.5em minus 0.4em\relax {MIT} Press, 1998.

\bibitem{TS32.298}
3GPP, ``{Technical Specification Group Services and Systems Aspects; Charging
  management; Charging Data Record (CDR) parameter description (Release 13)},''
  3GPP, TS 32.298, v13.3.0, 2016.

\bibitem{TS32.401}
------, ``Telecommunication management; performance management (pm); concept
  and requirements (release 10),'' 3GPP, TS 32.401, v10.2.0, 2010.

\bibitem{IETFRFC7012}
{Internet Engineering Task Force (IETF)}, ``{Information Model for IP Flow
  Information Export (IPFIX)},'' 3GPP, Tech. Rep. RFC 7012, 2013.

\bibitem{TS37.320}
3GPP, ``{Technical Specification Group TSG RAN; Radio measurement collection
  for Minimization of Drive Tests (MDT), Release 10},'' 3GPP, TS 37.320,
  v2.0.0, 2010.

\bibitem{D4D}
\BIBentryALTinterwordspacing
{Orange}. {The D4D Challenge}. [Online]. Available:
  \url{{http://www.d4d.orange.com/en/Accueil}}
\BIBentrySTDinterwordspacing

\bibitem{TS27.007}
3GPP, ``{Technical Specification Group Core Network and Terminals; AT command
  set for User Equipment (UE) (Release 12)},'' 3GPP, TS 27.007 v12.1.0, 2013.

\bibitem{opencellID}
\BIBentryALTinterwordspacing
{OpenCellID}. {Largest Open Database of Cell Towers and Geolocation}. [Online].
  Available: \url{{https://opencellid.org/}}
\BIBentrySTDinterwordspacing

\bibitem{opensignal}
\BIBentryALTinterwordspacing
{OpenSignal}. {3G and 4G {LTE} Cell Coverage Map}. [Online]. Available:
  \url{{https://opensignal.com/}}
\BIBentrySTDinterwordspacing

\bibitem{antenasgsm}
\BIBentryALTinterwordspacing
{Antenas GSM}. [Online]. Available: \url{{http://antenasgsm.com/}}
\BIBentrySTDinterwordspacing

\bibitem{googlegeo}
\BIBentryALTinterwordspacing
{Google geolocation api}. [Online]. Available:
  \url{{https://developers.google.com/maps/}}
\BIBentrySTDinterwordspacing

\bibitem{BuiOWL}
{N. Bui, J. Widmer}, ``{Owl: a reliable online watcher for {LTE} control
  channel measurements},'' \emph{arXiv preprint arXiv:1606.00202l}, 2016.

\bibitem{Trinh17}
{H. D. Trinh, N. Bui, J. Widmer, L. Giupponi, P. Dini}, ``{Analysis and
  Modeling of Mobile Traffic Using Real Traces},'' \emph{submitted to IEEE
  Personal Indoor and Mobile Radio Communications (PIMRC)}, 2017.

\bibitem{Peng13}
{M. Peng, D. Liang, Y. Wei, J. Li, and H. H. Chen}, ``Self-configuration and
  self-optimization in {LTE}-advanced heterogeneous networks,'' \emph{IEEE
  Communications Magazine}, vol.~51, no.~5, pp. 36--45, 2013.

\bibitem{stephenMLB}
{S. S. Mwanje, A. Mitschele-Thiel}, ``{Minimizing Handover Performance
  Degradation Due to {LTE} Self Organized Mobility Load Balancing},''
  \emph{77th IEEE Conference on Transactions on Vehicular Technology (VTC
  Spring)}, pp. 1--5, 2013.

\bibitem{munoz2}
{P. Mu\~{n}oz, R. Barco, I. de la Bandera}, ``{Optimization of load balancing
  using fuzzy {Q-Learning} for next generation wireless networks},''
  \emph{Expert Systems with Applications, Elsevier}, vol.~40, pp. 984--994,
  2013.

\bibitem{mlb}
{P. Mu\~{n}oz, R. Barto, I. de la Bandera}, ``{Load Balancing and Handover
  joint optimization in {LTE} networks using Fuzzy logic Reinforcement
  Learning},'' \emph{Computer Networks Elsevier}, 2014.

\bibitem{junishi}
{J. Suga, Y. Kojima, M. Okuda}, ``{Centralized Mobility Load Balancing Scheme
  in {LTE} Systems},'' \emph{8th International Symposium on Wireless
  Communication Systems, Aachen}, 2011.

\bibitem{emil}
{E. Bergner}, ``{Unsupervised Learning of Traffic Patterns in Self-Optimizing
  4th Generation Mobile Networks},'' \emph{Master of Science Thesis, KTH
  Computer Science and Communications, Stockholm, Sweden}, 2012.

\bibitem{Franco15}
{C. A. S. Franco, J. R. B. de Marca}, ``{Load balancing in self-organized
  heterogeneous {LTE} networks: A statistical learning approach},'' \emph{IEEE
  Latin-American Conference on Communications (LATINCOM)}, pp. 1--5, 2015.

\bibitem{qin}
{W. Qin, Y. Teng, M. Song, Y. Zhang, and X. Wang}, ``{A {Q-learning} approach
  for mobility robustness optimization in {LTE}-{SON}},'' \emph{15th IEEE
  International Conference on, Communication Technology (ICCT)}, pp. 818--822,
  2013.

\bibitem{stephenMLB2}
{S. S. Mwanje, A. Mitschele-Thiel}, ``{Distributed cooperative {Q-learning} for
  mobility-sensitive handover optimization in {LTE} {SON}},'' \emph{IEEE
  Symposium the Computers and Communication (ISCC)}, pp. 1--6, 2014.

\bibitem{HOpablo}
{P. Mu\~{n}oz, R. Barco, and I. de la Bandera}, ``{On the Potential of Handover
  Parameter Optimization for Self-Organizing Networks},'' \emph{IEEE
  Transactions on Vehicular Technology}, 2013.

\bibitem{mro}
{N. Sinclair, D. Harle, I. A. Glover, J. Irvine, R. C. Atkinson}, ``{Parameter
  Optimization for {LTE} Handover using an Advanced SOM Algorithm},''
  \emph{IEEE Conference on Transactions on Vehicular Technology (VTC)}, 2013.

\bibitem{mrosinclair}
{N. Sinclair, D. Harle, I. Glover, J. Irvine, and R. Atkinson}, ``{An advanced
  SOM algorithm applied to handover management within {LTE}},'' \emph{IEEE
  Transactions on Vehicular Technology}, pp. 183--1894, 2013.

\bibitem{Farooq17}
{H. Farooq, A. Imran}, ``{Spatio-temporal Mobility Prediction in Proactive
  Self-Organizing Cellular Networks},'' \emph{IEEE Communications Letters},
  vol.~21, no.~2, pp. 370--373, 2017.

\bibitem{Ali16}
{Z. Ali, N. Baldo, J. Mangues-Bafalluy, and L. Giupponi}, ``Machine learning
  based handover management for improved qoe in {LTE},'' \emph{IEEE/IFIP
  Network Operations and Management Symposium}, pp. 794--798, 2016.

\bibitem{Ostlin04}
{E. Ostlin, H. J. Zepernick, H. Suzuki}, ``{Macrocell radio wave propagation
  prediction using an artificial neural network},'' \emph{IEEE Vehicular
  Technology Conference (VTC Fall)}, p.~57, 2004.

\bibitem{Quintero04}
{A. Quintero and O. Garc\'{i}a}, ``{A profile-based strategy for managing user
  mobility in third-generation mobile systems},'' \emph{IEEE Communications
  Magazine}, pp. 134--139, 2004.

\bibitem{Majumdar05}
{K. Majumdar and N. Das}, ``{Mobile user tracking using a hybrid neural
  network},'' \emph{in Wireless Networks}, pp. 275--284, 2005.

\bibitem{RAZAVI}
{R. Razavi, S. Klein, H. Claussen}, ``A fuzzy reinforcement learning approach
  for self-optimization of coverage in {LTE} networks,'' \emph{Bell Labs
  Tachnical Journal}, vol.~15, pp. 153--175, 2010.

\bibitem{naseer}
{M. Naseer ul Islam, A. Mitschele-Thiel}, ``{A Cooperative Fuzzy {Q-learning}
  for Self-Organized Coverage and Capacity Optimization},'' \emph{IEEE 23rd
  International Symposium on Personal Indoor and Mobile Radio Communications
  (PIMRC)}, Sept 2012.

\bibitem{jingyu}
{J. Li, J. Zeng, X. Su, W. Luo, J. Wang}, ``{Self-Optimization of Coverage and
  Capacity in {LTE} Networks Based on Central Control and Decentralized Fuzzy
  {Q-Learning}},'' \emph{International Journal of Distributed Sensor Networks,
  Hindawi}, 2012.

\bibitem{ccoEURASIP}
{S. Fan, H. Tian, and C. Sengul}, ``{Self-optimization of coverage and capacity
  based on a fuzzy neural network with cooperative reinforcement learning},''
  \emph{Journal on Wireless Communications and Networking}, 2014.

\bibitem{Galindo}
{A. Galindo-Serrano, L. Giupponi, G. Auer}, ``{Distributed Learning in
  Multiuser {OFDMA} Femtocell Networks},'' \emph{73rd IEEE Vehicular Technology
  Conference (VTC Spring)}, pp. 1--6, 2011.

\bibitem{dirani}
{M. Dirani and Z. Altman}, ``{A cooperative reinforcement learning approach for
  inter-cell interference coordination in {OFDMA} cellular networks},''
  \emph{8th Int Modeling and Optimization in Mobile, Ad Hoc and Wireless
  Networks (WiOpt)}, pp. 170--176, 2010.

\bibitem{blascoICIC}
{M. Bennis, S. M. Perlaza, P. Blasco}, ``{Self-Organization in Small Cell
  Networks: A Reinforcement Learning Approach},'' \emph{IEEE Transactions on
  Wireless Communications}, vol.~12, no.~7, 2013.

\bibitem{simsek}
{M. Simsek, A. Czylwik, A. Galindo-Serrano, L. Giupponi}, ``{Improved
  decentralized {Q-learning} algorithm for interference reduction in
  {LTE}-femtocells},'' \emph{5th International Workshop on Self-Organizing
  Networks IWSON, Glasgow}, 2015.

\bibitem{miozzo}
{M. Miozzo, L. Giupponi, M. Rossi, P. Dini}, ``{Distributed {Q-Learning} for
  Energy Harvesting Heterogeneous Networks},'' \emph{IEEE ICC 2015 Workshop on
  Green Communications and Networks with Energy Harvesting, Smart Grids, and
  Renewable Energies}, June 2015.

\bibitem{annaCAMAD}
{A. Dudnikova, P. Dini, L. Giupponi, D. Panno}, ``{Multiple Criteria plus Fuzzy
  Logic Switch Off Method for Dense Heterogeneous Networks},'' \emph{20th IEEE
  International Workshop on Computer Aided Modelling and Design of
  Communication Links and Networks, Guildford (UK)}, 2015.

\bibitem{es1}
{E. Ternon, P. Agyapong, L. Hu, A. Dekorsy}, ``{Energy Savings in Heterogeneous
  Networks with Clustered Small Cell Deployments},'' \emph{11th IEEE
  International Symposium on Wireless Communications Systems (ISWCS)}, 2014.

\bibitem{es2}
{E. Ternon, P. Agyapong, L. Hu, and A. Dekorsy}, ``{Database-aided Energy
  Savings in Next Generation Dual Connectivity Heterogeneous Networks},''
  \emph{IEEE WCNC'14 Track 3: Mobile and Wireless Networks}, 2014.

\bibitem{jmoysenSH}
{J. Moysen and L. Giupponi}, ``{A Reinforcement Learning based solution for
  Self-Healing in {LTE} networks},'' \emph{80th IEEE Vehicular Technology
  Conference (VTC Fall), Vancouver, Canada}, 2014.

\bibitem{onireti}
{O. Onireti, A. Zoha, J. Moysen, A. Imran, L. Giupponi, M. Ali Imran, A.
  Abu-Dayya}, ``{A Cell Outage Management Framework for Dense Heterogeneous
  Networks},'' \emph{IEEE Transactions on Vehicular Technology}, vol.~64, pp.
  2097--2113, April 2015.

\bibitem{ARSALAN}
{A. Saeed, O. G. Aliu, M. A. Imran}, ``{Controlling Self Healing Cellular
  Networks using Fuzzy Logic},'' \emph{{IEE Wireless Communications and
  Networking Conference (WCNC), Paris, France}}, April 2012.

\bibitem{fedor}
{F. Chernogorov, J. Turkka, T. Ristaniemi, A. Averbuch}, ``{Detection of
  Sleeping Cells in {LTE} Networks Using Diffusion Maps},'' \emph{IEEE 73rd
  Vehicular Technology Conference (VTC Spring)}, pp. 1--5, may 2011.

\bibitem{khabib}
{E. J. Khatib, R. Barco, A. Gomez-Andrades, I. Serrano}, ``{Diagnosis based on
  genetic fuzzy algorithms for {LTE} Self-Healing},'' \emph{IEEE Transactions
  on Vehicular Technology}, 2015.

\bibitem{RANA}
{R. M. Khanafer, B. Solana, J. Triola, R. Barco, L. Moltsen, Z. Altman, P.
  Lazaro}, ``{Automated Diagnosis for UMTS Networks Using Bayesian Network
  Approach},'' \emph{IEEE Transactions on Vehicular Technology}, vol.~57, 2008.

\bibitem{emcoc}
{G. F. Ciocarlie, S. N. Sanneckaczki}, ``{Detecting Anomalies in Cellular
  Networks Using an Ensemble Method},'' \emph{9th CNSM and Workshops}, 2013.

\bibitem{Alias16}
{A. Multazamah, S. Navrati, R. Abhishek}, ``{Efficient Cell Outage Detection in
  5G Het-Nets Using Hidden Markov Model},'' \emph{IEEE Communications Letters},
  vol.~20, no.~3, 2016.

\bibitem{Xue14}
{W. Xue, M. Peng, Y. Ma, and H. Zhang}, ``Classification-based approach for
  cell outage detection in self-healing heterogeneous networks,'' \emph{IEEE
  Wireless Communications and Networking Conference (WCNC)}, p. 2822–2826,
  2014.

\bibitem{Zoha14}
{A. Zoha, A. Saeed, A. Imran, M. A. Imran, and A. Abu-Dayya}, ``A {SON}
  solution for sleeping cell detection using low-dimensional embedding of mdt
  measurements,'' \emph{IEEE 25th Annual International Symposium on Personal,
  Indoor, and Mobile Radio Communication (PIMRC)}, pp. 1626--1630, 2014.

\bibitem{Chernov14}
{S. Chernov, F. Chernogorov, D. Petrov, T. Ristaniemi}, ``Data mining framework
  for random access failure detection in {LTE} networks,'' \emph{25th IEEE
  Annual International Symposium on Personal, Indoor, and Mobile Radio
  Communication (PIMRC)}, pp. 1321--1326, 2014.

\bibitem{Barco05}
{R. Barco, V. Wille, and L. D\'{i}ez}, ``{System for automated diagnosis in
  cellular networks based on performance indicators},'' \emph{European
  Transactions on Telecommunications}, pp. 399--409, 2005.

\bibitem{Khanafer08}
{R. M. Khanafer, B. Solana, J. Triola, R. Barco, L. Moltsen, Z. Altman, P.
  Lazaro}, ``{Automated diagnosis for UMTS networks using Bayesian network
  approach},'' \emph{IEEE Transactions on Vehicular Technology}, vol.~57, p.
  2451–2461, 2008.

\bibitem{hafiz}
{H. Y. Lateef, A. Imran, A. Abu-dayya}, ``{A Framework for Classification of
  Self-Organising Network Conflicts and Coordination Algorithms},'' \emph{IEEE
  Personal Indoor and Mobile Radio Communications PIMRC}, 2013.

\bibitem{Berna_coord}
{O. Iacoboaeia, B. Sayrac, S. Jemaa, P. Bianchi}, ``{{SON} Coordination for
  parameter conflict resolution: A reinforcement learning framework},''
  \emph{IEEE Wireless Communications and Networking Conference 2014 (IEEE WCNC
  2014)}, April 2014.

\bibitem{lbHo}
{P. Mu\~{n}oz, R. Barco, and S. Flores}, ``{Conflict Resolution Between Load
  Balancing and Handover Optimization in {LTE} Networks},'' \emph{IEEE
  Communications Letters}, vol.~18, pp. 1795--1798, 2014.

\bibitem{jmoysenEurasip}
{J. Moysen and L. Giupponi}, ``{Self-coordination of parameter conflicts in
  {D-SON} architectures: a Markov decision process framework},'' \emph{EURASIP
  Journal on Wireless Communications and Networking}, 2015.

\bibitem{chernogorov12}
{F. Chernogorov and T. Nihtil{\"{a}}}, ``{QoS Verification for Minimization of
  Drive Tests in {LTE} Networks},'' \emph{75th {IEEE} Vehicular Technology
  Conference (VTC Spring)}, pp. 6--9, may 2012.

\bibitem{chernogorov13}
{F. Chernogorov and J. Puttonen}, ``{User satisfaction classification for
  Minimization of Drive Tests QoS verification},'' \emph{24th {IEEE} Personal
  Indoor and Mobile Radio Communications ({PIMRC}), London, United Kingdom},
  pp. 2165--2169, September 2013.

\bibitem{jmoysenISCC}
{J. Moysen, L. Giupponi, J. Mangues-Bafalluy}, ``{On the Potential of Ensemble
  Regression Techniques for Future Mobile Network Planning},'' \emph{21th IEEE
  Symposium on Computers and Communications (ISCC), Messina, Italy}, 2016.

\bibitem{jmoysenPIMRC}
------, ``{A Machine Learning enabled network Planning tool},'' \emph{27th
  {IEEE} Personal Indoor and Mobile Radio Communications (PIMRC), Valencia,
  Spain}, 2016.

\bibitem{balint}
{B. Dar\'{o}czy, A. Bencz\'{u}r, P. Vaderna}, ``{Machine Learning based session
  drop prediction in {LTE} networks and its {SON} aspects},'' \emph{5th
  International Workshop on Self-Organizing Networks (IWSON), Glasgow}, 2015.

\bibitem{turkka3}
{J. Turkka, D. Gil}, ``{Anomaly Detection Framework for Tracing Problems in
  Radio Networks},'' \emph{10th International Conference on Networks (ICN)},
  2011.

\bibitem{policy}
{R. Romeikat, B. Bauer, T. Bandh, G. Carle, H. Sanneck, L. C. Schmelz},
  ``{Policy-driven workflows for mobile network management automation},''
  \emph{6th International Wireless Communications and Mobile Computing
  Conference (IWCMC)}, pp. 1111--1115, 2010.

\bibitem{matiasCCO}
{V. Buenestado, M. Toril, S. Luna, J. M. Ruiz, A. Mendo}, ``{Self-tuning of
  remote electrical tilts based on call traces for coverage and capacity
  optimization in {LTE}},'' \emph{IEEE Transactions on Vehicular Technology},
  2017.

\bibitem{matias}
{M. Toril, R. Acedo-Hern\`{a}ndez, S. Luna-Ram\'{i}rez, A. S\'{a}nchez, C.
  \'{U}beda}, ``{Automatic clustering algorithms for indoor site selection in
  {LTE}},'' \emph{Hindawi Mobile Information Systems}, 2017.

\bibitem{AnaMariaMDT}
{A. Galindo-Serrano, B. Sayrac, S. Ben Jemaa, J. Riihij\"{a}rvi and P.
  M\"{a}h\"{o}nen}, ``{Harvesting MDT Data: Radio Environment Maps for Coverage
  Analysis in Cellular Networks},'' \emph{International Conference on Cognitive
  Radio Oriented Wireless Networks (CROWNCOM)}, 2013.

\bibitem{Clark13}
{D. Clark, et all.}, ``{A Knowledge plane for the Internet},'' \emph{in proc.
  of the 2003 Conference on Applications, Technologies, Architectures, and
  Protocols for Computer Communications}, 2003.

\bibitem{Rios2013}
{L. M. Rios, N. V. Sahinidis}, ``Derivative free optimization: A review of
  algorithms and comparison of software implementations,'' \emph{Journal of
  Global Optimization}, pp. 1247--1293, 2013.

\bibitem{Derbel2009}
{H. Derbel, N. Agoulmine, M. Salan}, ``Anema: Autonomic network architecture to
  support self-configuration and self-optimization in ip networks,''
  \emph{Computer Networks}, pp. 418--430, 2009.

\bibitem{Zorzi16}
e.~a. M.~Zorzi, ``{COBANETS}: A new paradigm for cognitive communications
  systems,'' \emph{International Conference on Computing, Networking and
  Communications {ICNC}}, 2016.

\bibitem{Mestres16}
{A. Mestres, et all.}, ``{Knowledge Defined Networking},'' \emph{arXiv preprint
  arXiv:1606.06222.v2}, 2016.

\bibitem{Yahia17}
{I. Ben Yahia, et all.}, ``{CogNitive 5G networks: Comprehensive operator use
  cases with machine learning for management operations},'' \emph{20th
  Conference on Innovations in Clouds, Internet and Networks (ICIN)}, 2017.

\bibitem{Bendriss17}
{J. Bendriss, et all.}, ``{AI for SLA Management in Programmable Networks},''
  \emph{International Conference on Design of Reliable Communication Networks},
  2017.

\bibitem{european}
{F. Ricciato, P. Widhalm, M. Craglia and F. Pantisano}, ``{Estimating
  population density distribution from network-based mobile phone data},
  journal={JRC and AIT},'' 2015.

\bibitem{laner}
{M. Laner, P. Svoboda, P. Romirer, N. Nikaein, F. Ricciato}, ``{A Comparison
  Between One-Way Delays in Operating HSPA and {LTE} Networks}, journal ={8th
  Int'l workshop on Wireless Network Measurements (WINMEE'12), Pedeborn,
  Germany},'' 2012.

\bibitem{lena}
\BIBentryALTinterwordspacing
{Centre Tecnol\`{o}gic de Telecomunicacions de Catalunya (CTTC)}. {The LENA-EPC
  Network simulator}. [Online]. Available:
  \url{{http://networks.cttc.es/mobile-networks/software-tools/lena/}}
\BIBentrySTDinterwordspacing

\bibitem{enisa}
{European Union Agency for Network and Information Security (enisa)},
  ``{Privacy by design in big data},'' \emph{{An overview of privacy enhancing
  technologies in the era of big data analytics}}, 2015.

\bibitem{Khairulliza16}
{K. A. Sallehab and L. Janczewski}, ``{Technological, organizational and
  environmental security and privacy issues of big data: A literature
  review},'' \emph{Procedia Computer Science, ELSEVIER}, pp. 19--28, 2016.

\bibitem{Bengio13}
{Y. Bengio, A. Courville, P. Vincent}, ``{Representation Learning: A Review and
  New Perspectives},'' \emph{IEEE Transactions on Pattern Analysis and Machine
  Intelligence}, vol.~35, no.~8, pp. 1798--1828, 2013.

\bibitem{DL2016}
{I. Goodfellow, Y. Bengio, A. Courville}, \emph{{Deep Learning}}.\hskip 1em
  plus 0.5em minus 0.4em\relax MIT Press, 2016,
  \url{http://www.deeplearningbook.org}.

\bibitem{nashQ}
{J. Hu and M. P. Wellman}, ``{Nash {Q-Learning} for General-Sum Stochastic
  Games},'' \emph{Journal of Machine Learning Research}, vol.~4, pp.
  1039--1069, 2003.

\bibitem{zorzi}
{M. Zorzi, A. Zanella, A. Testolin, M. Filipo de Grazia, M. Zorzi},
  ``{Cognition-Based Networks: A New Perspective on Network Optimization Using
  Learning and Distributed Intelligence},'' \emph{IEEE Access The journal for
  rapid open access publishing, Special Section on Artificial Intelligence
  Enabled Networking}, vol.~3, pp. 1512--1530, 2015.

\end{thebibliography}

\end{document}